\documentclass[twocolumn,tighten,usenames,dvipsnames]{aastex62}
\usepackage[normalem]{ulem}
\usepackage{verbatim}
\usepackage{amssymb}
\usepackage{amsmath}
\usepackage{xspace}

\newcommand{\pt}[1]{{\color{magenta} #1}}

\def\risco{r_{\rm ISCO}}
\newcommand{\themis}{{\sc Themis}\xspace}

\newcommand{\vrtt}{VRT$^2$\xspace}

\shorttitle{Spacetime tomography}
\shortauthors{Tiede et al.}

\begin{document}
\title{Spacetime Tomography Using The Event Horizon Telescope}

\author[0000-0003-3826-5648]{Paul Tiede}
\affiliation{Department of Physics and Astronomy, University of Waterloo, 200 University Avenue West, Waterloo, ON, N2L 3G1, Canada}
\affiliation{ Waterloo Centre for Astrophysics, University of Waterloo, Waterloo, ON N2L 3G1 Canada}
\affiliation{Perimeter Institute for Theoretical Physics, 31 Caroline Street North, Waterloo, ON, N2L 2Y5, Canada}

\email{ptiede@perimeterinstitute.ca}

\author[0000-0001-9270-8812]{Hung-Yi Pu}
\affiliation{Perimeter Institute for Theoretical Physics, 31 Caroline Street North, Waterloo, ON, N2L 2Y5, Canada}
\email{hpu@perimeterinstitute.ca}

\author[0000-0002-3351-760X]{Avery E. Broderick}
\affiliation{Department of Physics and Astronomy, University of Waterloo, 200 University Avenue West, Waterloo, ON, N2L 3G1, Canada}
\affiliation{ Waterloo Centre for Astrophysics, University of Waterloo, Waterloo, ON N2L 3G1 Canada}
\affiliation{Perimeter Institute for Theoretical Physics, 31 Caroline Street North, Waterloo, ON, N2L 2Y5, Canada}

\author[0000-0003-2492-1966]{Roman Gold}
\affiliation{CP3-Origins, University of Southern Denmark, 
Campusvej 55, DK-5230 Odense M, Denmark}
\affiliation{ Institut f\"ur Theoretische Physik, Goethe-Universit\"at Frankfurt, Max-von-Laue-Stra{\ss}e 1, D-60438 Frankfurt am Main, Germany}
\affiliation{Perimeter Institute for Theoretical Physics, 31 Caroline Street North, Waterloo, ON, N2L 2Y5, Canada}

\author[0000-0001-7387-9333]{Mansour Karami}
\affiliation{Department of Physics and Astronomy, University of Waterloo, 200 University Avenue West, Waterloo, ON, N2L 3G1, Canada}
\affiliation{Perimeter Institute for Theoretical Physics, 31 Caroline Street North, Waterloo, ON, N2L 2Y5, Canada}

\author[0000-0002-4146-0113]{Jorge A. Preciado-L\'opez}
\affiliation{Perimeter Institute for Theoretical Physics, 31 Caroline Street North, Waterloo, ON, N2L 2Y5, Canada}

\begin{abstract}
  \indent We have now entered the new era of high-resolution imaging astronomy with the beginning of the Event Horizon Telescope (EHT). The EHT can resolve the dynamics of matter in the immediate vicinity around black holes at and below the horizon scale.  One of the candidate black holes, Sagittarius A*, flares 1--4 times a day depending on the wavelength.  A possible interpretation of these flares could be hotspots generated through magnetic reconnection events in the accretion flow.  In this paper, we construct a semi-analytical model for hotspots that include the effects of shearing as a spot moves along the accretion flow. We then explore the ability of the EHT to recover these hotspots.  Even including significant systematic uncertainties, such as thermal noise, diffractive scattering, and background emission due to an accretion disk, we were able to recover the hotspots and spacetime structure to sub-percent precision.  Moreover, by observing multiple flaring events we show how the EHT could be used to \textit{tomographically map spacetime}.  This provides new avenues for testing relativistic fluid dynamics and general relativity near the event horizon of supermassive black holes.  
\end{abstract}

\keywords{black hole physics --- Galaxy: center --- methods: phenomenological models --- methods: numerical --- accretion, accretion disks --- galaxies: jets --- radiative transfer -- submillimeter: general}


\section{Introduction}
\label{sec:intro}
The Event Horizon Telescope (EHT), has an unparalleled resolution of $20\mu\rm as$, providing the ability to resolve the event horizon's of at least two black holes \citep{Doeleman2008,2009astro2010S..68D,EHTCI,EHTCII,EHTCIII}.  This is accomplished by using very long-baseline interferometry (VLBI) with stations spread across the Earth.  Previous studies of Sagittarius A* (Sgr A*) have already provided constraints on the size of the central object \citep{Fish2011,Fish2016,Ortiz-Le2016,Johnson:2015} and the nature of its accretion flow \cite{Broderick2011,Broderick2016} and imaged the central ring feature of M87 \cite{EHTCI,EHTCIV,EHTCV,EHTCVI}.

One of the primary science goals of the EHT is to probe accretion and spacetime in the strong-gravity regime.  Many of these methods involve using time averaged measurements by creating static images \citep{Akiyama2017,Broderick2014,Chan2015,Johannsen2016}. Time averaging, however, creates a degeneracy between when and where emission arises near the black hole. By exploring time variability, we can break this degeneracy and explore the dynamics of the emission region and structure of spacetime.

While M87 is static over a day, Sgr A* displays strong variability through flaring events \citep{Genzel2003} that can often (especially for bright X-ray flares) be seen (almost) simultaneously across multiple bands  \citep{Fazio2018}, from sub-millimeter \citep{Fish2011}, to infrared \citep{Gillessen2006,Witzel2012,Witzel2018} and X-ray \citep{Neilsen2013,Ponti}. This emission appears to come from a compact region near the innermost stable circular orbit (ISCO) of the black hole and is presumed to be from dynamical structures within the accretion flow \citep{Marrone2006,Gillessen2006,GravityHS}. An explanation for these flares comes from the creation of localized ``hotspots'' of non-thermal electrons in the accretion disk surrounding the black hole and has been proposed by several authors: \citet{BroderickLoeb2005,BroderickLoeb2006, Eckart2006Pol} and previously \citet{Dovciak2004}.
A natural origin is magnetic-reconnection events within the accretion disk analogous to solar flares, an unavoidable consequence of radiatively inefficient accretion models. 

Previous work on modeling orbiting hotspots assumed compact spherical Gaussian structures that remain coherent during its orbit\citep{BroderickLoeb2005, BroderickLoeb2006}. In general, however, hotspots are expected to be embedded within a differentially rotating accretion disk, and therefore will shear and expand.  Furthermore, shearing due the differential flow of an accretion disk can lead to large observational differences for NIR flares \citep{Eckart2008a,Eckart2009,ZamaninasabShear}. For the EHT, including shearing may be imperative since it is sensitive to horizon scale physics.  To address these concerns, we developed a computationally efficient model including generic shear and expansion, while ensuring that spot number density is locally conserved.

As hotspots orbit, they probe different parts of spacetime. By observing a flare, we not only probe the null structure of spacetime from the emission but also how massive matter evolves in the vicinity of the event horizon.  As we will show below, observing a single hotspot with the EHT may lead to high precision spin measurements. Furthermore, since each hotspot will form at different radii, every flare will probe different regions of spacetime. Combining multiple flares would then amount to constructing a tomographical map of spacetime, leading to a new test of GR in the vicinity of black holes.

In practice, recovering the hotspots from EHT observations could be difficult. The effective beam size of the EHT is $13 \mu\mathrm{as}$, meaning that the hotspot may not be sufficiently resolved by the EHT to precisely probe spacetime.  Furthermore, there are several important systematics present for Sgr A*, such as scattering and the background accretion flow.
To address these questions we used \themis \citep{ThemisCode}, a Bayesian parameter estimation framework designed for use with the EHT. \themis was designed to recover the posterior for models applicable to EHT observations and was used extensively in the first EHT results on M87 \cite{EHTCI, EHTCV, EHTCVI}. Therefore, we will numerically explore the ability of the EHT to perform inference on hotspots using synthetic data that matches the configuration of the EHT 2017 array and address the impact of some potential systematics.

This paper is organized as follows. In Section~\ref{sec:model} we present an original hotspot model that incorporates shearing and expansion while conserving particle number. Section~\ref{sec:ehtMovies}, explores the ability of the EHT in the 2017 configuration, to extract a hotspot from potential observations of Sgr A*. As a result, we analyze whether the differential flow parameters such as angular velocity and black hole spin, are degenerate. Additionally, we study how the background flow, scattering, and the accretion disk inclination relative to our line of sight impact our results. Section~\ref{sec:tomo} details how multiple hotspots can be used to tomographically map spacetime using the EHT and constructs hypothetical maps using the EHT.

\begin{figure*}[!t]
\centering
\includegraphics[width=\textwidth]{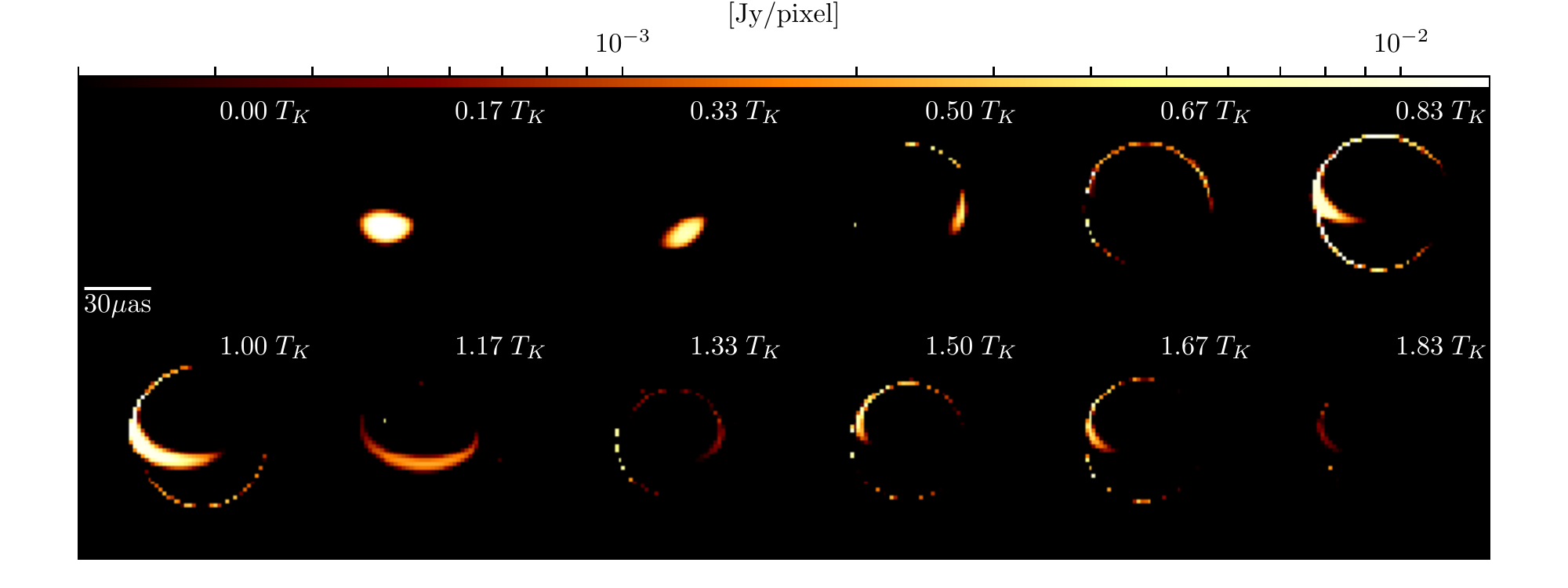}
\caption{Frames of a shearing spot movie intensity map in log-scale, with $64\times 64$ image resolution set around Sgr A*. The movie lasts for two Keplerian orbits at a radius of $5.23M$, and contains 12 frames in total. The parameters are $a_*=0.5\; \cos\Theta=0.5,~n_0=5.5\times 10^7,\; R_s=0.5M,\; r_0=5.23M\simeq1.25~\risco,\;\phi_0=-90^\circ,\; \alpha=0.05,\; \kappa=0.95$.}\label{fig:spot_movie}
\end{figure*}

\section{Hotspot model}\label{sec:model}

Hotspot models have been used to explain flares and variability in Sgr A* observations using coherent Gaussian hotspots \cite{BroderickLoeb2005, BroderickLoeb2006}. Later models allowed for adiabatic expansion \citep{Eckart2009} and shearing \citep{ZamaninasabShear}. However, in the latter work, the emission was restricted to a 2D disk and radiative transfer effects were ignored. Here, we describe a model that includes both shearing and expansion of a hotspot along a stationary and axisymmetric velocity field. Furthermore, this model includes the 3-dimensional structure of the hotspot and includes the effects of radiative transfer as will be described below. Additionally, we describe a semi-analytical procedure for evolving the hotspot density that is marginally more computationally expensive than the coherent hotspot model.

\subsection{Density profile evolution}

In \citet{BroderickLoeb2005,BroderickLoeb2006}, hotspots are modeled by orbiting, symmetric Gaussian electron overdensities.  The orbital position of the center of the spot, $y^\mu_0$ is determined by integrating the accretion flow four-vector field $u^\mu$ around which the hotspot number density is given by,
\begin{equation}\label{eq:OrbittingSpotProfile}
  n_e(y^\mu) = n_0e^{-\left(\Delta r^\mu\Delta r_\mu + (\Delta r^\mu u_{\mu})^2\right)/(2R_s^2)},
\end{equation}
where $\Delta r^{\mu}=y^\mu-y^\mu_0$, is the displacement vector from the center of the Gaussian spot, $R_s$ the spot size, and $u_\mu$ is evaluated at the spot center.  While this model is computationally efficient, it ignores the potential for differential motion within the spot.  For a Keplerian velocity field, this approximation is rapidly violated: the inner edge of the spot has advanced relative to the outer-edge by one radian in $r/3\pi R_s$ orbits; within $1/3\pi$ orbits it has advanced by $R_s$.

Therefore, we will assume that a hotspot will travel passively on some specified (e.g. background accretion flow) velocity field $u^\mu$. That is, the density is described by the continuity equation coupled with the condition that the particles move with the background flow:
\begin{align}
  \nabla_\mu (n_e u^\mu) &= 0,\label{eq:ParticleConservation}\\ 
  \frac{d x^\mu}{d \tau} &= u^\mu.\label{eq:vectorFlow}
\end{align}
We will denote the solution family of Equation \eqref{eq:vectorFlow} by the map $\varphi_\tau(y^\mu)$. By fixing $y^\mu$ we can consider $x^\mu(\tau)=\varphi_{\tau}(y^\mu)$ as describing the motion of a particle at some time $\tau$. Namely, $x^\mu(\tau)$ satisfies Equation~\eqref{eq:vectorFlow}, with initial condition $x^\mu(\tau=0)=y^\mu$. On the other hand, if we consider $\tau$ as fixed then $\phi(y^\mu) = \varphi_\tau(y^\mu)$ describes the coordinates of a family of observers at rest with the background flow. Therefore, we can say for small $\tau$ that $\varphi_\tau(y^\mu)$ forms a 1-parameter family of diffeomorphisms. Therefore, we can solve Equation \eqref{eq:ParticleConservation} using the method of characteristics, giving
\begin{align}
  u^\mu\partial_\mu n_e + n_e\nabla_\mu u^\mu = \frac{d}{d\tau} n_e + n_e \nabla_\mu u^\mu = 0
\end{align}
and
\begin{equation}\label{eq:spot_density_gen}
  n_e(\tau,x^\mu) = n_{e,0}(y^\mu) \exp\left(-\int_{\tau_0}^\tau \nabla_\mu u^\mu d \tau\right),
\end{equation}
where $y^\mu = \varphi^{-1}_{\tau}(x^\mu)$ is the initial position of the spot and $n_{e,0}$ the initial proper density profile.
We can simplify this further by noting
\begin{equation}
  \nabla_\mu u^\mu = \frac{1}{\sqrt{-g}} \partial_\mu(\sqrt{-g}u^\mu),
\end{equation}
and thus,
\begin{equation}\label{eq:spotevolution}
  n_e(\tau,x^\mu) = n_{e,0}(y^\mu)\frac{\sqrt{-g(y^\mu)}}{\sqrt{-g(x^\mu)}}\exp\left(-\int_{\tau_0}^\tau \partial_\mu u^\mu d\tau\right).
\end{equation}
Interpreting this result physically, we see that there are two forms of expansion included in the model. The first is spacetime expansion and is encoded by the ratio of the metrics. The second is the expansion of the velocity field itself irrespective of the background spacetime. Note that for a hotspot outside the innermost stable circular orbit (ISCO) traveling on a Keplerian orbit, Equation~\eqref{eq:spot_density} simplifies to $n_e(\tau,x^\mu)=n_{e,0}(y^\mu)$, since $g(x^\mu)$ is constant and $\partial_\mu u^\mu$ vanishes. This does not mean that there is no deformation, as $y^\mu = \varphi^{-1}_\tau(x^\mu)$. Instead, the only deformation is due to shearing which does not affect the proper density\footnote{We define shearing as the symmetric traceless part of the tensor $\nabla_\mu u_\nu$. Since it is trace-free does not directly impact the proper density of the hotspot (see Equation \eqref{eq:spotevolution}).}. Note however, that this is an idealization and in reality the hotspot will adiabatically expand during its evolution, meaning that its proper density will decrease over time.

In this paper, we will focus on hotspots orbiting in the equatorial place, although the density extends outside. Furthermore, we will assume that the spot has negligible vertical motion compared to radial and azimuthal motion. To describe our vector field we follow \cite{Pu2016}. A form of the accretion flow that obeys the restrictions mentioned above can be parameterized by
\begin{equation}
  u^\mu = (u^t, u^r, 0, u^t\Omega),\qquad \Omega = u^\phi/u^t.
\end{equation}
The normalization condition $u^\mu u_\mu=-1$ for a black hole metric in Boyer-Lindquist like coordinates gives
\begin{equation}
  u^t = \sqrt{\frac{1 + g_{rr}(u^r)^2}{-g_{tt} - 2\Omega g_{t\phi} - \Omega^2g_{\phi\phi} }}.
\end{equation}
From this we can see that we require that $g_{tt} + 2\Omega g_{t\phi} + \Omega^2g_{\phi\phi} < 0$. To specify $u^r$ and $\Omega$ we will use a combination of Keplerian and free-fall motion. Namely,
\begin{align}
  u^r &= u^r_{\rm K} + \alpha(u^r_{\rm ff}-u^r_{\rm K})\label{eq:radialAFV},\\
  \Omega &= \Omega_{\rm K} + (1-\kappa)(\Omega_{\rm ff}-\Omega_{\rm K})\label{eq:angAFV},
\end{align}
where $\alpha,~\kappa\in[0,1]$, are two free-parameters that control the rate of free-fall and the sub-Keplerian motion respectively. Note that our definition of $\alpha$ differs from that of \cite{Pu2016}. For the Keplerian component, outside the ISCO, $u^r_{\rm K}=0$. Inside the ISCO $u^r_K\neq0$ and is specified by matching the energy and angular momentum at the ISCO. This choice of velocity field brackets a useful collection of accretion flows. For example, taking $\alpha=0,~\kappa=1$  gives a Keplerian orbit and $\alpha=0,\kappa=0$ free-fall motion.

Due to the fact that this vector field is independent of the coordinates $t$ and $\phi$, we have that $\partial_\mu u^\mu = \partial_r u^r$. Therefore, $\partial_\mu u^\mu = (u^r)^{-1}\dot{u^r}$, and Equation~\eqref{eq:spot_density_gen} simplifies to
\begin{equation}\label{eq:spot_density}
  n_e(\tau,x^\mu) = n_{e,0}(y^\mu)\frac{\sqrt{-g(y^\mu)}}{\sqrt{-g(x^\mu)}}
\frac{u^{r}(y^{\mu})}{u^{r}(x^{\mu})}
  \;.
\end{equation}
This semi-analytic formula greatly increases the computational speed of the hotspot, and is the same order of computational complexity as the coherent spot used in \cite{BroderickLoeb2005}.

As a final note, in principle, any smooth function could be used for the initial density profile. However, in this paper, we will assume that the spot is initially given by Equation~\eqref{eq:OrbittingSpotProfile}. There are two reasons for this.  First, this profile allows us to compare the hotspot evolution results to \cite{BroderickLoeb2005, BroderickLoeb2006}.  Second, since any image will be distorted by the interstellar scattering screen \citep{Bower2006, Johnson2018}, whose diffractive or blurring component is effectively a Gaussian with a semi-major axis of $22\mu\rm as$ most small-scale structure of the hotspot will be unresolved. Therefore, since we are assuming that a hotspot forms from local microphysics, i.e. fast magnetic reconnection, we expect it to be contiguous, and the initial profile can be approximated as a Gaussian.
 
\subsection{Radiative transfer and Ray-tracing}
To create hotspot models that can be compared to EHT data, we need to create realistic images. This means that near the black hole, general relativistic (GR) and radiative transfer effects need to be included. These effects include the geometric and gravitational time delays across the source (often called ``slow light'') and the strong gravitational lensing that magnifies the emission region and produces secondary images associated with photons that complete half orbits around the black hole. Furthermore, optical depth becomes important near the black hole since material moving towards the detector will have an increased apparent density due to Doppler effects, making it optically thick.

To incorporate these effects we use the covariant ray-tracing and radiative transfer code \vrtt (vacuum ray-tracing radiative transfer). For the EHT observation band (230GHz) we assume the hotspot spectra in the plasma rest frame is given by the synchrotron self-absorption model from \citet{Broderick2004}, with a local plasma energy spectral index of $s=2.25$. At the observing frequency of the EHT ($\sim 230$GHz), synchrotron cooling processes will be sub-dominant to the shearing timescale. The shearing timescale is roughly the orbital period of the hotspot around Sgr A*. Taking the hotspot to be around the ISCO, we get $t_{\rm shear}\sim 10-30$~min for a $4\times 10^6~M_\odot$ black hole. The timescale for synchrotron cooling can be estimated from $t_{\rm synch} \sim 3\times10^7 \nu_9^{-0.5} B^{-3/2}$\,s, where $\nu_9$ is the frequency in GHz and $B$ is in Gauss. Taking $B\sim 10-50$~G we find that  $t_{\rm synch} \sim 1-20$~hours at 230~Ghz. Therefore, in this paper we ignore the cooling break and evolution during the hotspots orbit. Note that in the other bands, e.g., the near-infrared and X-ray, cooling and inverse Compton effects likely become important \citep{Fazio2018} and will need to be included. 

To model the magnetic field assumed to arise from an accretion disk, we followed \citet{Broderick2016} and used a toroidal magnetic field with a fixed plasma beta set to $10$. Below we will also consider what happens when the hotspot is embedded in an accretion flow. In this case, we use the accretion flow model from \citet{Broderick2016}. This model includes thermal electrons set in a radial power law in density and temperature, where the power-law indices for the thermal electron and temperature distribution are $-1.1, -0.84$ respectively. Furthermore, we also include non-thermal electrons with a radial power-law index $-2.02$ for the number density, and $1.24$ for the photon spectral index. Note that these parameters were chosen to match the values in \citet{Broderick2016}, which are the best fit values to the spectrum of Sgr A*. 

The optical depth of the hotspot depends on a number of quantities, such as the proper density of the hotspot, its orbital parameters, and the orientation of the orbital plane relative to our line of sight. For instance, a hotspot near the ISCO at $230$~GHz with an inclination angle of $60^\circ$, will tend appear to be optically thick when it is moving towards us due to Doppler beaming and when moving away will be optically thin. Furthermore, as the hotspot shears its effective optical depth will change. All of these effects are automatically included into the relativistic radiative transfer that occurs in the construction of each frame of the movie.

This completely describes the shearing hotspot model used in this paper. In summary, this model, ignoring any spectral information, requires 10 parameters to describe the evolution which are:
\begin{enumerate}
  \item spin parameter $a_*$.
  \item cosine of the inclination, $\cos\Theta$ of the black hole relative to the image screen.
  \item Spot electron density $n_0$ in Equation \eqref{eq:OrbittingSpotProfile}.
  \item Spot characteristic size, $R_s$, in units of $M$\pt{\footnote{We are using geometrical units, where $G=c=1$ in this paper.}} from Equation \eqref{eq:OrbittingSpotProfile}.
  \item Spot injection time, $t_0$, in units of $M$ for an observer: the time which the spot is instantaneously injected into the accretion flow. The actual spot appears at a fixed proper time for an observer in a locally flat co-moving frame with $u^\mu$. This means for the observers time, the hotspot gradually starts to appear.\footnote{Since we assumed that hotspots are created from local microphysics, i.e. from fast magnetic reconnection, we expect the hotspot to appear suddenly and be localized initially.}
  \item Initial hotspot radius, $r_0$, i.e. the position of the spot center in Boyer–Lindquist coordinates when it is initially injected.
  \item Initial hotspot azimuthal angle, $\phi_0$, i.e. the angle in Boyer-Lindquist coordinates of the hotspot center when it is injected into the accretion flow.
  \item Radial accretion flow parameter, $\alpha$, in Equation \eqref{eq:radialAFV}.
  \item Angular accretion flow parameter, $\kappa$, in Equation \eqref{eq:angAFV}.
  \item Position angle of black hole spin and orbital axis, $\xi$. 
\end{enumerate}

\autoref{fig:spot_movie} presents a 12 frame movie with the parameters $a_*=0.5,\; \cos\Theta=0.5,\; n_0=5.5\times 10^7,\; R_s=0.5,\; t_0=-6M,\; r_0=5.23M\simeq 1.25 \risco,\; \phi_0=-90^\circ,\; \alpha=0.05,\; \kappa=0.99,\; \xi=0^\circ$. The initial time of the spot was chosen to be $-6M$ due to time delay effects from ray-tracing. The inclination was chosen to be equal to the expected inclination for a uniform distribution on a sphere, which is close to the observed inclination found in \citet{Broderick2016}. Different inclinations will be explored in Section~\ref{sec:inclination}. Additionally, we chose the initial azimuthal angle of the hotspot to be $-90^\circ$ to ensure that when the spot first passes in front of the black hole relative to our line of sight it hasn't appreciably sheared. The initial radius was chosen to be close to the ISCO since this is where a hotspot would be expected to be found, which was recently seen in \cite{GravityHS}. Furthermore, an exploration of how radius and spin effects the images will be discussed below. The accretion flow parameters $\alpha$ and $\kappa$ were chosen to be close to a perfect Keplerian motion since we anticipate this will be the motion of the accretion disk in Sgr A*. The radial size of the spot $R_s$ was chosen to be $0.5M$ to test the ability of the EHT to resolve spots similar to the beam size of the EHT. The total observation time was $2T_K$ at the initial spot location, where $T_K$ is the Keplerian orbital period and for a spot at $5.3\risco=53~\rm min$.   Each of the twelve frames has a resolution of $64\times 64$. While higher resolutions can be used, we made this choice for two reasons: First, it is low enough to allow for movies to make in a reasonable timescale for parameter estimation. Second, higher resolutions did not appreciably impact parameter estimation, which is presented in the next section.

\begin{figure*}[!ht]
\centering
\includegraphics[width=\linewidth]{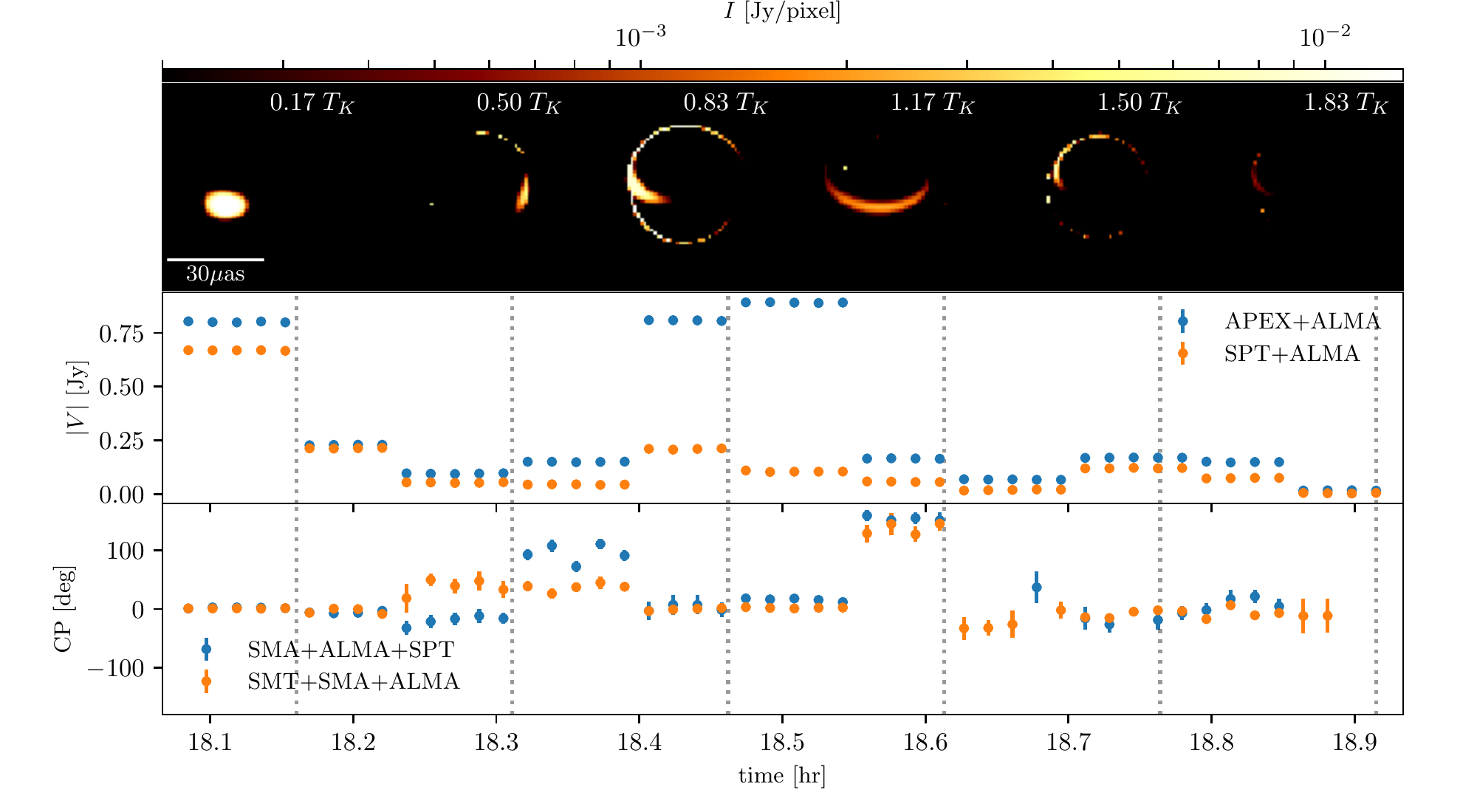}
\caption{VLBI observables for the twelve frame movie shown in \autoref{fig:spot_movie}. The top panels show a subset of the 12 frames in logarithmic scale. The middle panels show synthetic VM at the APEX+ALMA (blue) and SPT+ALMA (orange) for the 2017 EHT observation baselines as a function of time. The bottom panels show the CP as a function of time at the SMA+ALMA+SPT (blue), and SMT+SMA+ALMA (orange) triangles. The gray dotted lines show where the time of movie snapshots in the top panel are taken. Note that the discrete jumps in the observations are due to the fact that we only used a 12 frame movie.}\label{fig:spot_vmcp}
\end{figure*}

\section{Shearing hot spots with the EHT}\label{sec:ehtMovies}
To assess whether the EHT has sufficient fidelity to recover a shearing hotspot's parameters, we need to first convert our ray-traced movies, such as the one shown in \autoref{fig:spot_movie}, to the VLBI data that, e.g., the EHT 2017 array will observe. The EHT, like all VLBI arrays, measure not the intensity map of an image but instead quantities associated with its Fourier transform, namely the complex visibilities. Each pair of EHT stations form a baseline for which a complex visibility is recorded. In this section, we will describe the procedure we used to convert our images into EHT synthetic data. In the first part, we describe what observables the EHT measures in more detail, and how we create synthetic data for the movie shown in \autoref{fig:spot_movie}. In the second, we report a Bayesian MCMC parameter estimation exercise and analyze the EHT's ability to reconstruct shearing hot spot parameters from the simulated data. In the third part, we will analyze the impact of two systematics (diffractive scattering and a background RIAF) will have on the parameters posterior distribution. In all cases, we find that the EHT can recover all the true hotspot parameters to sub-percent precision at $95\%$ confidence about the median.

\subsection{Creating Synthetic EHT Data}\label{sec:ehtdata}
To explore how well the EHT can recover hotspots, we first need to convert the movie into interferometric EHT data. The EHT measures complex visibilities defined by, 
\begin{equation}\label{eq:vis}
  V_{ij} = \int d\alpha d\beta\; I(\alpha,\beta)e^{2\pi i (\alpha u + \beta v)}.
\end{equation}
Due to phase calibration issues from atmospheric turbulence, the phases of individual stations are practically randomized. To get around this, visibility amplitudes (VA), $\left|V_{ij}\right|$, are used, for which calibration or gain uncertainties are typically around 10\% and can effectively be modeled (see \cite{ThemisCode}). To recover some information about the complex phase of the visibilities closure phases (CP) are constructed
\begin{equation}\label{eq:CP}
  \Phi_{i,j,k} = \arg(V_{ij}V_{jk}V_{ki}),
\end{equation}
which are just the sum of the phases of a triplet of visibilities. Because the baselines close, i.e., they form a triangle $(u,v)_{ij} + (u,v)_{jk} + (u,v)_{ki} = 0$, all station-specific gain errors vanish from the CPs. 

In this paper, we will use VA and CP data to explore the ability of the EHT 2017 to reconstruct shearing hotspot parameters at 230GHz. To convert spot intensity maps into interferometric data we use the \textit{Event Horizon Telescope Imaging Library} (\texttt{eht-imaging})\footnote{\url{https://github.com/achael/eht-imaging}}; \citep{ehtim2016,ehtim2018,ehtim2019}). \texttt{eht-imaging} provides the ability to convert intensity maps into VA and CP data that uses the EHT 2017 array configuration including the correct baseline information, atmospheric thermal noise, and gain errors. This allows us to directly sample the image at the baselines the EHT will sample in a given observation window. We created an observation at $51544 \rm MJD$ starting at $1800\rm hr$ with scan and integration time of $61\rm s$ and $31 \rm s$ respectively. The total observation was $53~\rm min$, which corresponds to two Keplerian orbits for a spot at $5.25M$. The specific baselines we used are shown by the white points in \autoref{fig:spot_riaf_sys}. We also include Gaussian thermal noise in all observations, where the error, $\sigma_{ij}$, for the $(i,j)$ baseline is determined by 
\begin{equation}
  \sigma_{ij} = \frac{1}{0.88}\sqrt{\frac{\mathrm{SEFD}_i\mathrm{SEFD}_j}{2\,t_{\rm int}\nu_{\mathrm{bw}}}},
\end{equation}
where $\nu_{\rm bw}$ is the bandwidth of the observation which we set to $4\times10^9\rm Hz$, $2\nu_{\rm bw}t_{\rm int}$ is the number of independent samples of the two station baseline (i,j),  and the $1/0.88$ factor is due to two-bit quantization \citep{TMS}. Furthermore, the SEFD (System Equivalent Flux Density) of each station are provided by \texttt{eht-imaging}. While gain errors can be included by \texttt{eht-imaging}, we will ignore their impact for simplicity.

When creating the synthetic VA and CPs we placed a signal to noise ratio (SNR) cut of 2 on every baseline and debiased the VA's according to 
\begin{equation}
\mathcal{V}_{ij} =\sqrt{|V_{ij}|^2 - \sigma_{ij}^2}.
\end{equation}
This allows us to approximate the error distributions of the VA and CP as a Gaussian to $<10 \%$ accuracy \citep{TMS, ThemisCode}. Furthermore, and SNR cut of 2 only removes a handful of measurements. The resulting visibility amplitudes and closure phases for a few baselines as a function of time are shown in \autoref{fig:spot_vmcp}.
Short baselines, which probe large scales, show modest variations associated with the hotspot, consistent with the light curve.  In contrast, long baselines, which probe small scales, exhibit large variations, associated with the rapidly varying structures within the image.  In both cases, the variations are easily identifiable, substantially exceeding the thermal noise.

\subsection{Extracting spacetime and spot parameters with \themis}\label{sec:param_estimation}
\begin{figure*}[!ht]
\centering
\includegraphics[width=\textwidth]{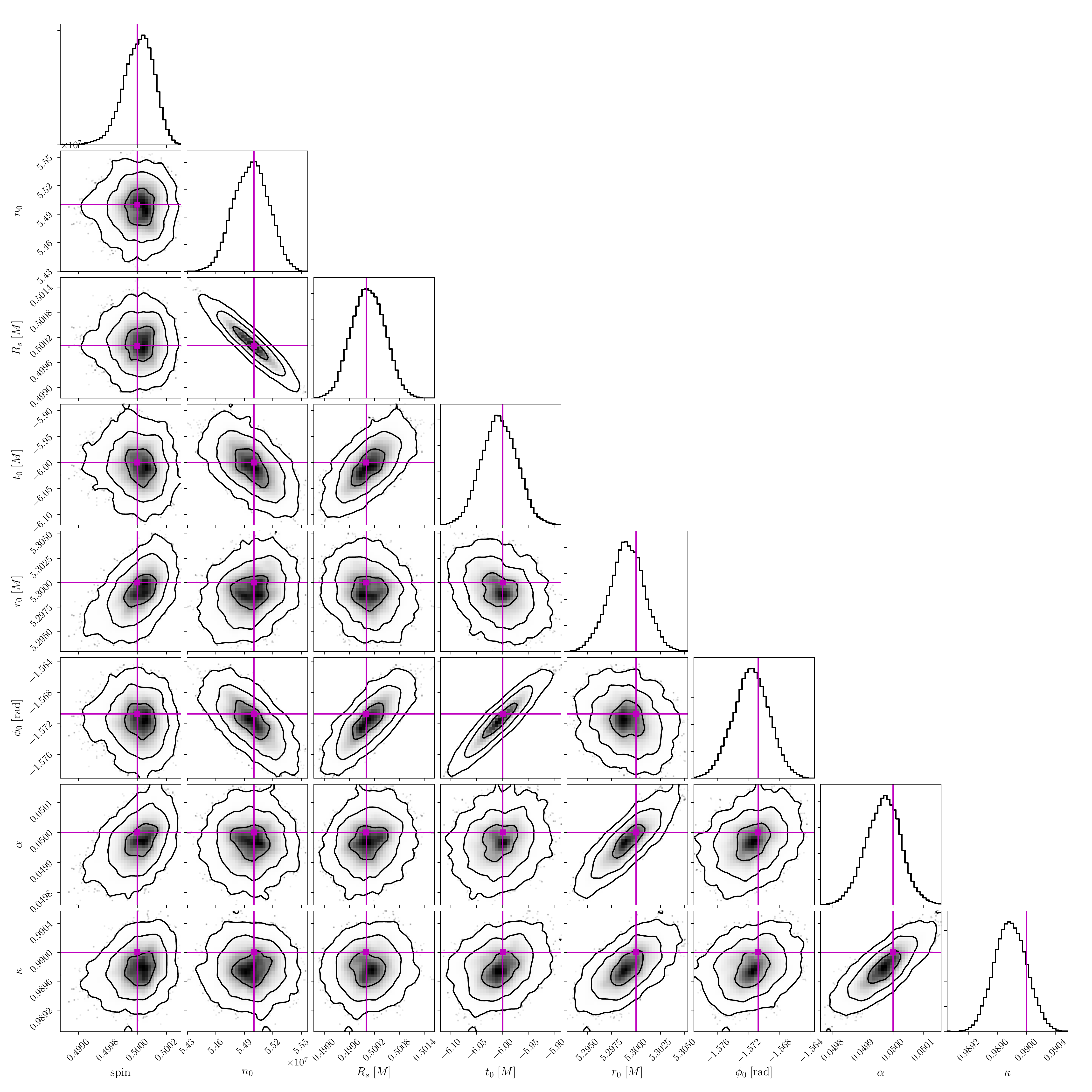}
\caption{Joint posterior probability distribution of shearing spot. The purple lines and points show the true values of the model, which is shown in \autoref{fig:spot_vmcp}.}\label{fig:spot_triangle}
\end{figure*}
To quantitatively examine the ability of the 2017 EHT array to recover and constrain hotspot parameters we use MCMC to recover the posterior distribution. To accomplish this, we used the software suite \themis. \themis is a highly extensible parameters estimation framework that was developed to deal with modeling and feature extraction of EHT observations. Furthermore, it can easily accommodate time variable structures. For more information about \themis see \citet{ThemisCode}. For modeling, we used \themis' non-marginalized Gaussian likelihoods for both the visibility amplitudes and closure phases.

We expect EHT observations during quiescent, non-flaring periods to place strong constraints on the orientation of the black hole spin and the images azimuthal orientation.  Therefore, we hold the black hole inclination, $\cos\Theta$, and image position angle, $\xi$, fixed to their ``true'' values during each MCMC run. The reasoning behind this is that when analyzing a real data set, we expect that imaging studies will provide a prior estimate for both parameters. In principle, these parameters could be allowed to vary, however, we found that it does not alter any of the results below. 

Flat priors where chosen for for $a_*\in (0,1)$, $R_s\in(0.01,10)$, $t_0\in(-200,200)$, $r_0\in(1.5M,20M)$, $\phi_0\in(-\pi,\pi)$, $\alpha\in(0,1)$, $\kappa\in(0,1)$, and $n_0$ used a logarithmic prior, ranging from $10^3$ to $10^{12}$. For the MCMC sampler we used the parallel tempered affine-invariant sampler originally detailed in \cite{Goodman2010,Vousden16}, with $48$ walkers, $6$ tempering levels, with temperature swaps every $50$ MCMC steps. To speed up convergence the walkers were started at the true values of the model. A single run for 1000 MCMC steps took $150\, 000$ core hours, on the Calcul Quebec and Compute Canada cluster Mp2.

\begin{figure*}[!ht]
\centering
\includegraphics[width=\textwidth]{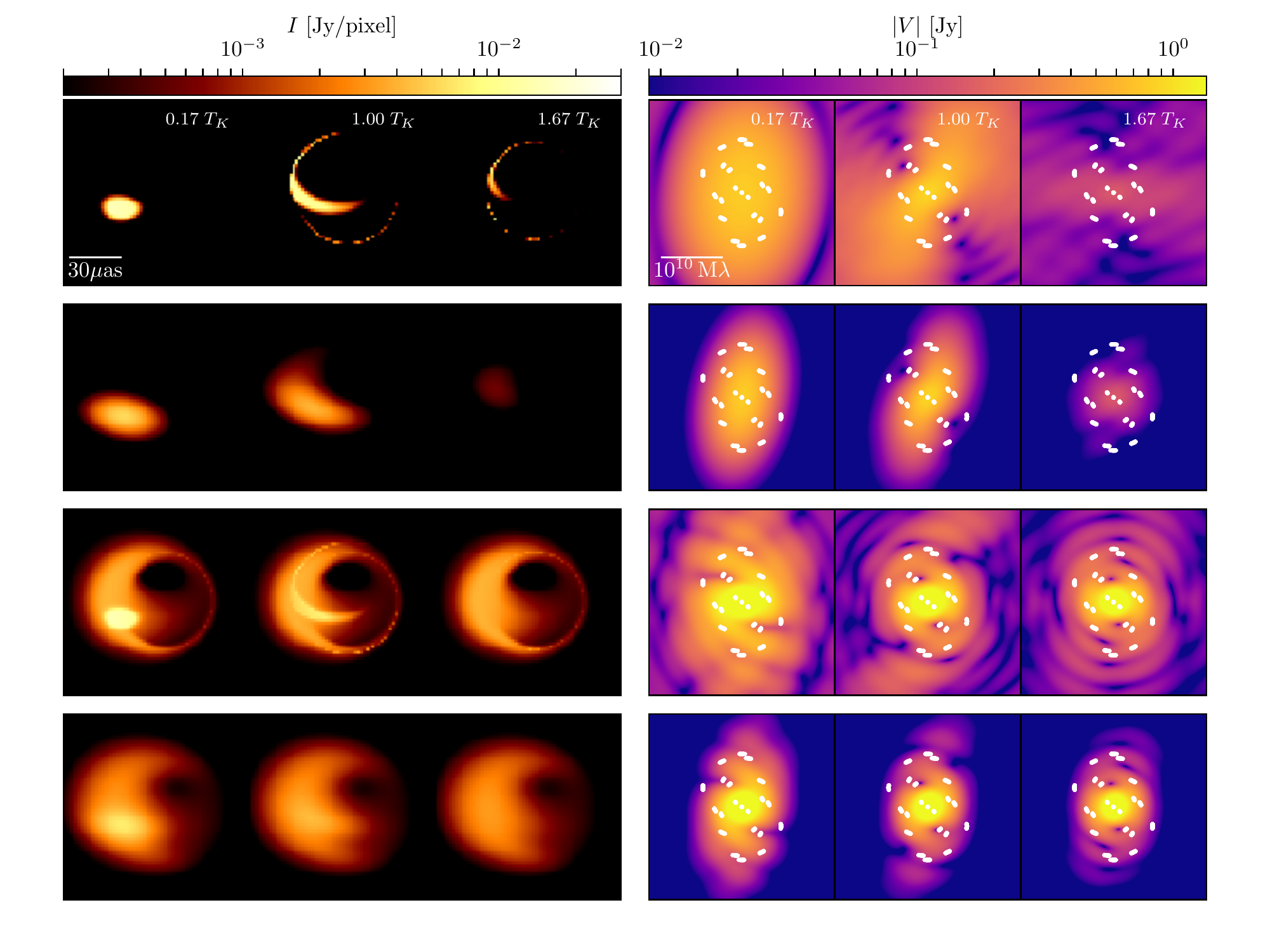}\\
\caption{Comparison of movies snapshots (left) and their corresponding visibility amplitudes (right) as different systematics are added. The white dots denote the (u,v) points sampled using the EHT array configuration described in the main text. The top figures are the base case with no systematics. The second from the top is with diffractive scattering, second from the bottom is with a RIAF and no scattering, and the bottom is with both a RIAF and scattering.}\label{fig:spot+riaf_pmap}
\end{figure*}

The joint parameter posterior probability distributions are shown in \autoref{fig:spot_triangle}. Every spot parameter is recovered with sub-percent accuracy. For example, the median spin and its $95\%$ confidence interval was $a_* =  0.50002^{+0.00017}_{-0.00023}$. Every marginalized distribution was single-peaked, showing no apparent degeneracies using the EHT 2017 array. The minimum reduced chi-square was found to be 1.0001 with 1640 degrees of freedom. Therefore, we conclude that the EHT 2017 array can accurately recover isolated shearing hotspots around Kerr black holes with high precision and accuracy. While these results are encouraging, we have ignored all potential systematics except the thermal noise present in the array. In the next section, we will study the impact of a few potential systematics that are important for Sgr A*.

\subsection{Adding Systematics -- Scattering and Background flows}\label{subsec:systematics}
In practice, the idealized observations described in the previous section are not directly applicable to EHT data, which are subject to a variety of additional systematic effects. Therefore here, we analyze how some of these systematics modify our conclusions. We focus on two systematics that are expected to dominate the error budget: diffractive scattering and a background accretion flow. Additionally, we analyze the impact of different inclination angles of the accretion disk on our parameter estimation.

\subsubsection{Diffractive Scattering}
Emission from SgrA* is scattered by interstellar electrons \citep{Bower2006, Johnson2018, Issaoun2019}.  This both blurs the image (diffractive scattering) and stochastically lenses the image (refractive scattering).  We consider the implications of the former for the reconstruction of shearing hot spots here, leaving the latter for future work. Diffractive scattering has the effect of washing out any structure below the scale of the blurring kernel. For our kernel, we use the empirically determined, wavelength-dependent asymmetric Gaussian kernel from \cite{Bower2006} with parameters,
\begin{equation}
\begin{aligned}\label{eq:scat_ellipse}
  \theta_{ \rm maj} &= 1.309\left(\frac{\lambda}{1\, \rm cm}\right)^2\,  \rm mas,\, \\
  \theta_{\rm min} &= 0.64\left(\frac{\lambda}{1\, \rm cm}\right)^2\,  \rm mas,\,
\end{aligned}
\end{equation}
where $\theta_{\rm maj,\rm min}$ is the FWHM of the semi-major/minor axis of the Gaussian scattering ellipse and $\psi=78^\circ$ its orientation. At $230 \mathrm{GHz}$ this corresponds to a semi-major axis FWHM of $22\mu\rm{as}$. The impact of diffractive scattering on the image is shown in the left panels of \autoref{fig:spot+riaf_pmap}. As expected, the diffractive scattering removes structure smaller than the typical kernel size of the image. For visibility amplitudes (right panels of \autoref{fig:spot+riaf_pmap}), this corresponds to the suppression of visibility amplitudes at high baselines length. The impact on the direct EHT observables for the hotspot movie in \autoref{fig:spot_movie} is shown in \autoref{fig:vlbi_sys}. Here we see that at long baselines, i.e. the SPT+ALMA baseline, variations in the VM are damped.  Note that since the scattering Kernel is a Gaussian the closure phases are not modified.

\subsubsection{RIAF Background}

\begin{figure}[!t]
\centering
\includegraphics[width=\linewidth]{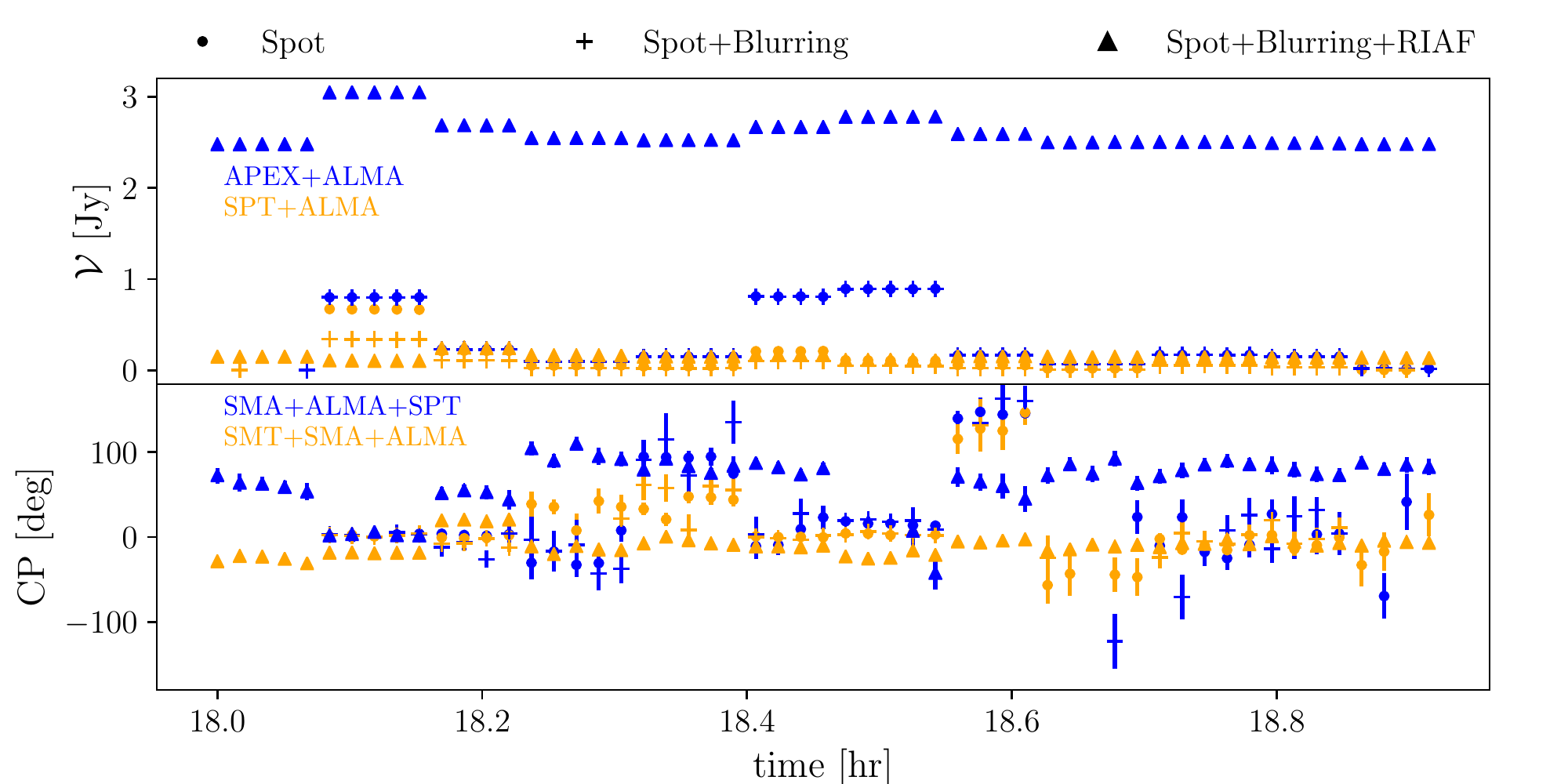}
\caption{Impact of scattering and a background RIAF on the VM (top) and CP (bottom) as the spot falls into the black hole. We again use the same 12 frame movie parameters as shown in \autoref{fig:spot_vmcp} }\label{fig:vlbi_sys}
\end{figure}

While hotspots can contribute substantially to the image flux, the main source of emission, the accretion disk, will typically dominate. Where it does not, its opacity will still obscure the hotspot emission. To include the impact of an accretion disk we included a radiatively inefficient accretion flow (RIAF) model from \cite{Broderick2016} fitted to past proto-EHT observations of Sgr A*. \autoref{fig:spot+riaf_pmap}, demonstrates how the RIAF background impacts hotspot movies. The impact is twofold. One, we see that regions, where the emission is very dim, is washed out by the background RIAF. Secondly, and most importantly, is the impact of optical depth from the background accretion flow. This effect is especially pronounced in the Doppler boosted region of the disk. In this region, the spot becomes entirely washed out after it passes through the ISCO. This suggests that hotspots appearing inside the ISCO will be much harder to observe with the EHT.


\begin{figure}[!t]
\centering
\includegraphics[width=\linewidth]{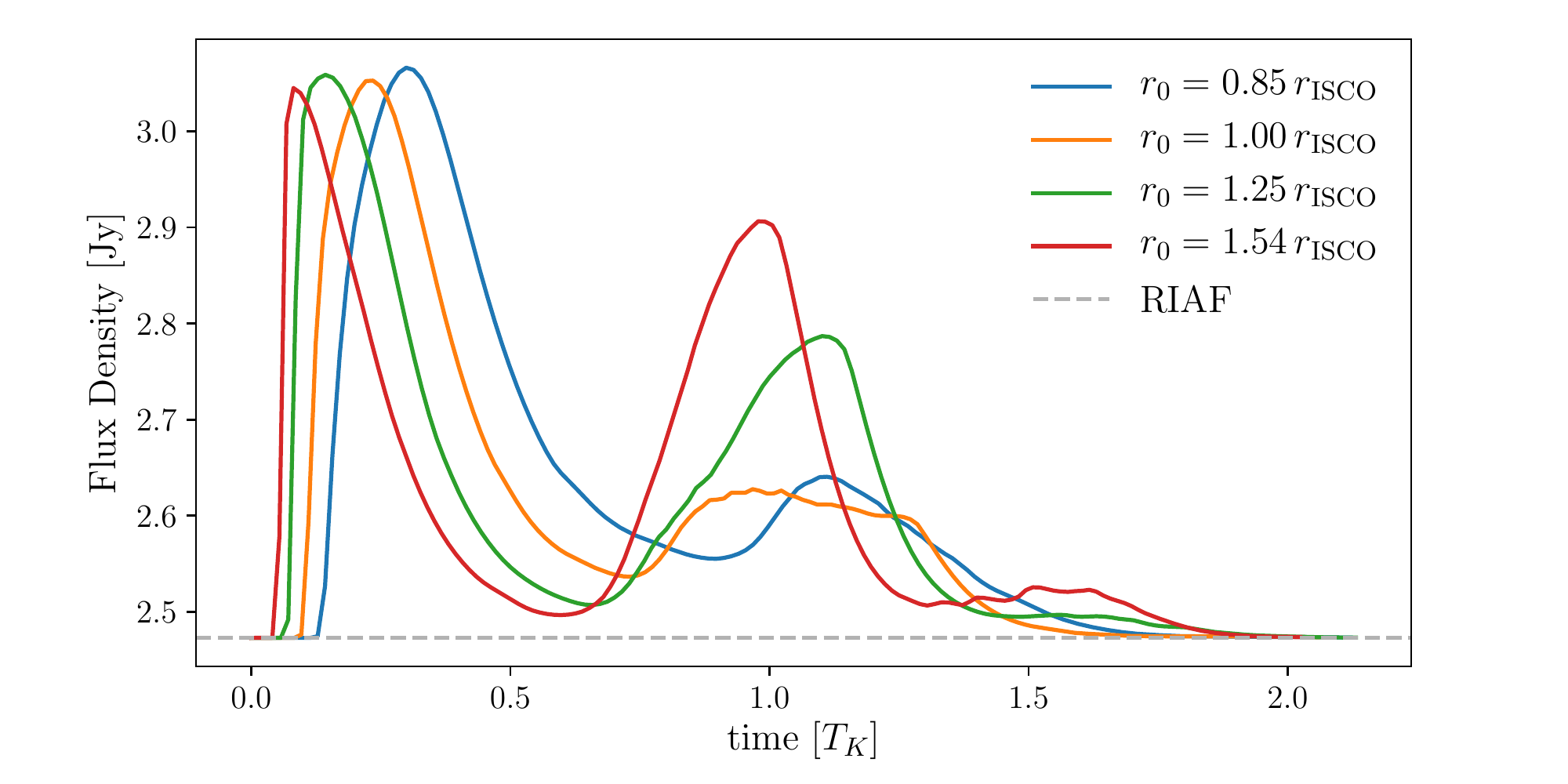}
\caption{VLBI lightcurves for a shearing spot with blurring and a RIAF background, around a black hole with spin parameter $a_*=0.5$, at the four different radii specified in Table~\ref{table:spot_experiment}. The units of the x-axis are given as a fraction for the Keplerian orbital period $T_K$ at the initial radius for the respective spot.}\label{fig:lightcurve}
\end{figure}
\begin{figure}[!t]
\centering
\includegraphics[width=\linewidth]{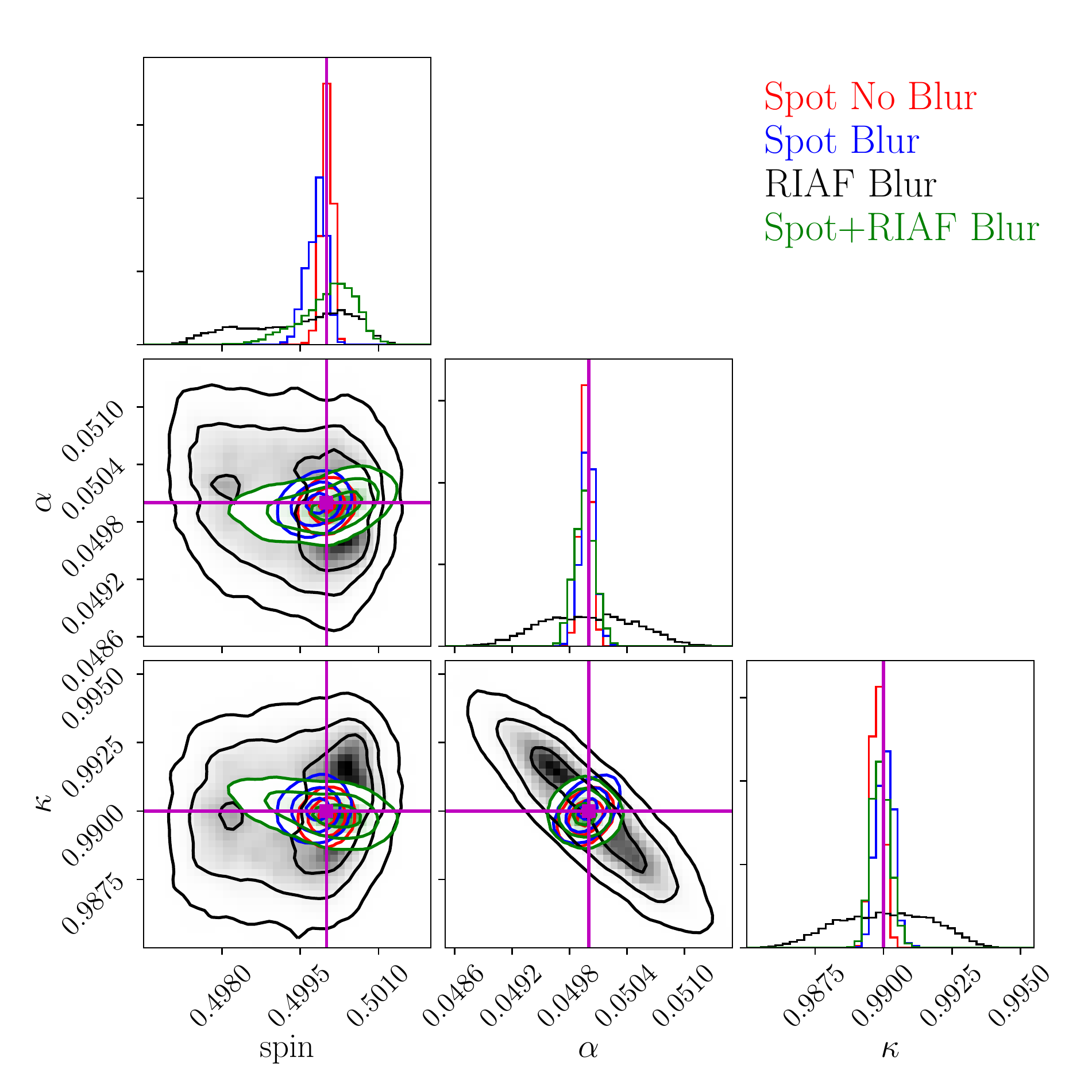}
\caption{Comparison of the joint-probability distribution of a shearing hotspot with different systematics. The black contours show represent the base model without diffractive scattering and a background RIAF. Blue the same spot but with diffractive scattering, and red if the same spot with diffractive scattering and a RIAF background. The model parameters are: spin of $0.5$ with a viewing inclination of $60^\circ$, accretion flow parameters are set of $\alpha=0.05$ and $\kappa=0.99$ (near-Keplerian), and the spot was initially placed at $1.25\cdot \risco=5.3\, M$ with a azimuthal angle of $-90^\circ$ and $\xi=0^\circ$.}\label{fig:spot_riaf_sys}
\end{figure}

\begin{figure*}[!ht]
  \centering
\includegraphics[width=\textwidth]{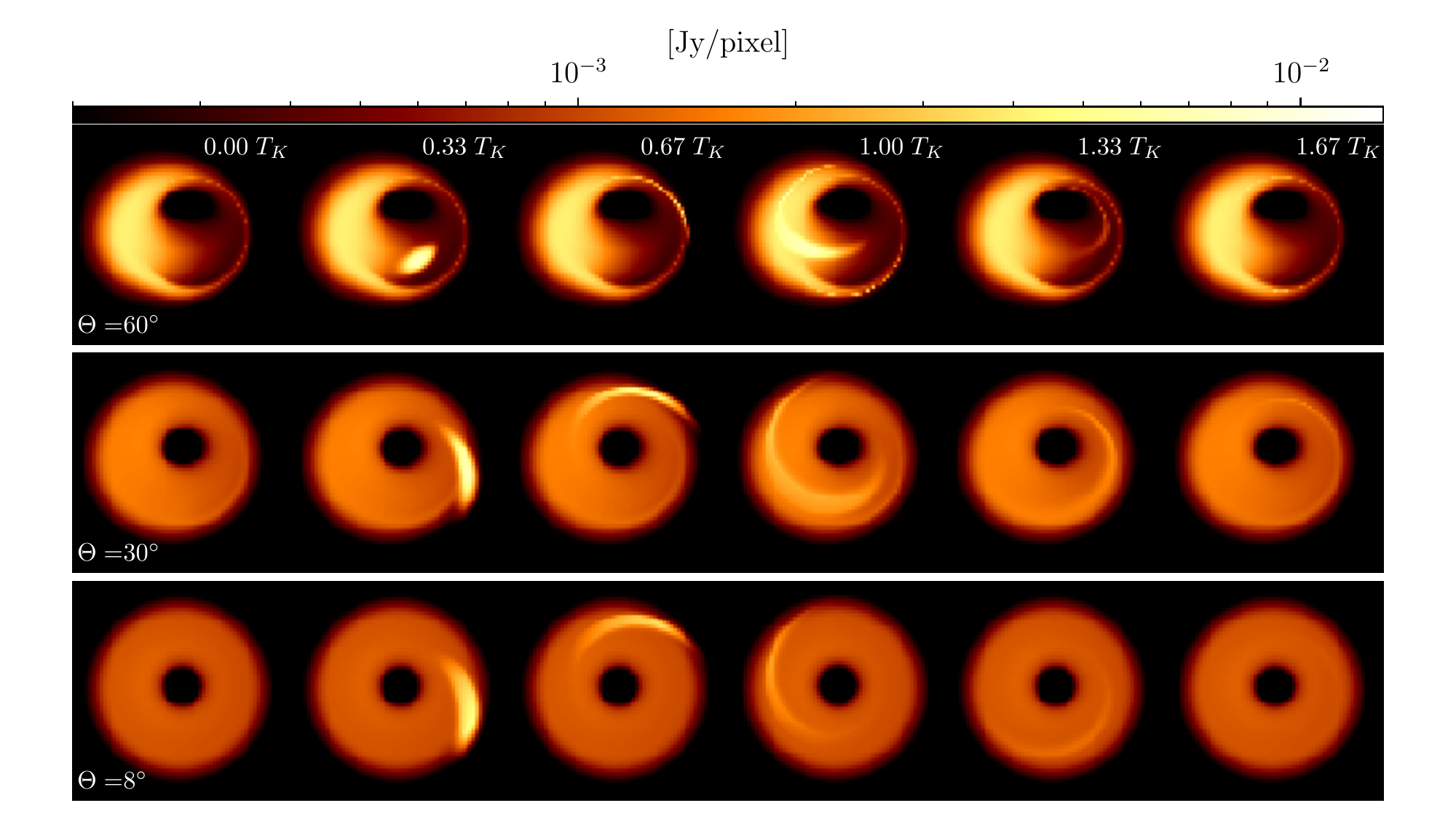}
\caption{Comparison of hotspot motion embedded in the \citet{Broderick2016} best fit RIAF model, over different accretion disk inclination angles $\Theta=60^\circ,\ 30^\circ,\ 8^\circ$. As the inclination angle becomes smaller lensing is suppressed since optical depth from the accretion disk becomes large.}\label{fig:movie_cos}
\end{figure*}
In terms of EHT observables, we see how the RIAF impacts the visibilities and images in Figures \ref{fig:spot+riaf_pmap} and \ref{fig:vlbi_sys}). After adding the RIAF, the spot brightness above the background drops from $0.8 \rm Jy$ to $0.5 \rm Jy$ from optical depth. The impact of optical depth is even more pronounced after the spot makes two complete orbits\footnote{Since the spot has radial motion it completes more than two orbits during the observation time.}, as is seen in the third panel of the third column of \autoref{fig:spot+riaf_pmap}. Before adding the RIAF, the hotspot extends across the entire face of the black hole and the secondary emission is visible. After, the hotspot is practically invisible.

Another way to see this impact is by analyzing how the light curve of the spot changes as the initial radius of the spot moves inwards, which is shown in \autoref{fig:lightcurve}. To assure a fair comparison as the initial radius of the hotspot is changed, we decrease the density constant $n_0$ too. This ensures the maximum brightness of the spot fixed to $\sim0.5 \rm Jy$.  Analyzing \autoref{fig:lightcurve} as the starting radius of the hotspot is moved inwards, it moves into the optically thick region sooner. This leads to the second orbit of the spot is increasingly obscured. As we will see below, for hotspots starting inside the ISCO, this negatively impacts the ability of the EHT to recover hotspots.

\subsubsection{Impact of systematics on parameter estimation}

To estimate the impact of the systematics on the parameter estimation, we use the procedure described in \autoref{sec:param_estimation}, with identical starting parameters, priors, and sampler options, but including the RIAF and blurring to the model when applicable. The impact of the systematics on the posterior distribution is shown in \autoref{fig:spot_riaf_sys}. Blurring does not appear to impact the posteriors substantially. One reason for this is that diffractive scattering doesn't change closure phases. Additionally, looking at \autoref{fig:spot+riaf_pmap}, blurring is a multiplication of the VA and can easily be inverted through modeling, since the kernel has no nulls in visibility space. However, when the background RIAF and diffractive scattering are both included the posteriors do broaden. For the black hole spin, the range of the inferred values increases by roughly a factor of two. However, we still have sub-percent precision, finding that $a_* = 0.5001^{+0.00073}_{-0.00129}$. Therefore, even with blurring and a RIAF background, the EHT 2017 array can recover hotspots and sub-horizon-scale physics to high accuracy.

In the absence of a background RIAF, all spacetime constraints arise from the shearing hot spot. In the presence of a RIAF, the morphology of the quiescent accretion flow provides additional information \citep{JohanssenII}. Thus, we seek to assess the improvement in the spin measurement, arising from the inclusion of the hot spot relative to the background RIAF. Namely, if the spots are improving the measurement of spin, we would expect the bound on the spin to improve relative to just fitting a RIAF background during the same observation. To test this, we used the same array configuration as for the hotspot simulations, and then created simulated data of a RIAF with the same parameters used previously. The results are shown by the black curves in \autoref{fig:spot_riaf_sys}, which compares the joint probability distribution for the spin and two accretion flow parameters. For the case of the RIAF, we find the spin is given by $a_* = 0.49964^{+0.00128}_{-0.00213}$. The $95\%$ errors are then $70\%$ larger than the hotspot and RIAF model. Therefore, we indeed see that catching a flare does improve EHT measurements of spin.

\begin{figure}[!h]
\centering
\includegraphics[width=\linewidth]{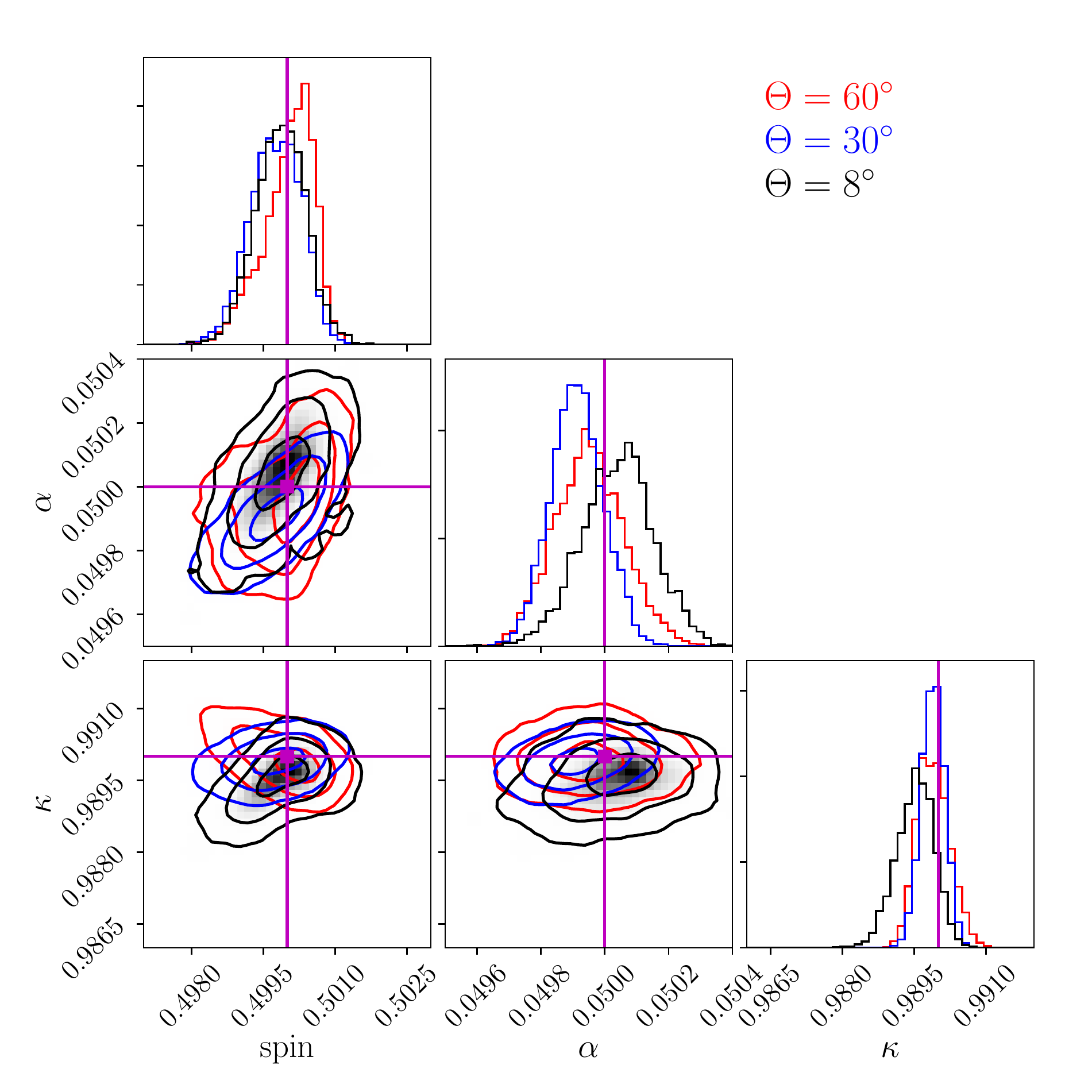}
\caption{Comparison of the joint-probability distribution of shearing hotspots with a background RIAF and diffractive scattering at different inclination angles. \textit{Red} represents the standard inclination angle used in this paper of $\Theta=60^\circ$ and is the same posterior as the green curve in \autoref{fig:spot_riaf_sys}. \textit{Blue} is for $\Theta=30^\circ$ and is roughly the inclination found by \citet{GravityHS}. \textit{Black} is for $\Theta=8^\circ$. In all instances the measurement of spin is similar to the standard $\Theta=60^\circ$ case.}\label{fig:triangle_cos}
\end{figure}

\begin{figure*}[!t]
\centering
\includegraphics[width=\linewidth]{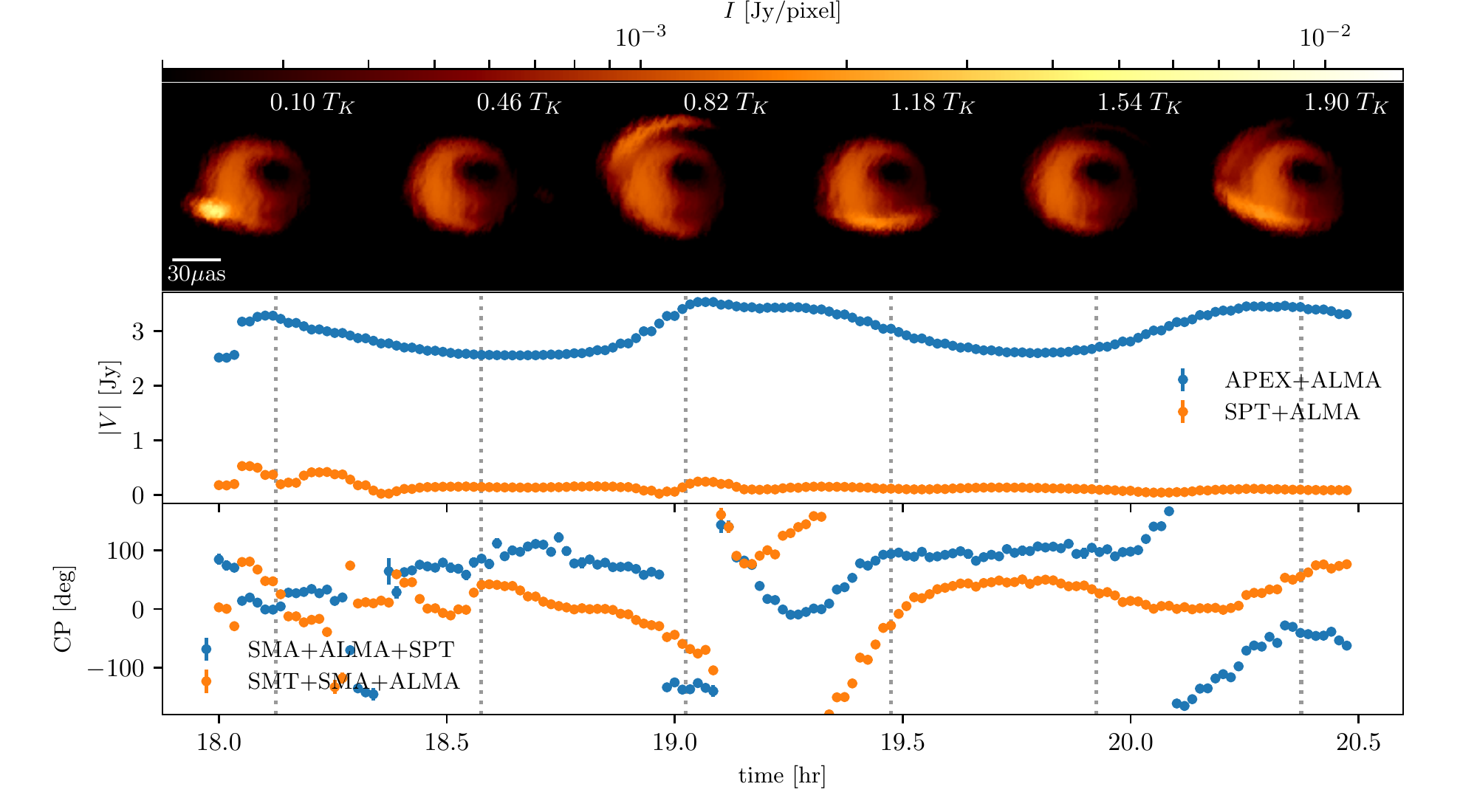}
\caption{Example of a spot movie with refractive scattering and a RIAF background included. The movies uses 100 frames over a $2.5 \rm hr$ window. The model parameters are: $a_*=0,\;  \cos\Theta=0.5,\; R_s=0.5,\; n_e=5.5\times 10^7,\; t_0=-6M,\; r_0=10M,\; \phi_0=-90^\circ,\; \alpha=0.05,\; \kappa=0.99,\; \xi=0^\circ$. The top panel shows a selection of frames of the movie the times specifed by the grey dotted lines. The middle and bottom panel show the VA and CP at a couple of baselines and triangles.}\label{fig:spot_ref}
\end{figure*}

\subsection{The impact of disk inclination}\label{sec:inclination}
  We have shown that the inclusion of a scattering screen and background accretion flow does not drastically alter our ability to extract hotspots. However, most of the conclusions so far have assumed that the inclination of the accretion disk is $\Theta = 60^\circ$. While this angle does match what \citet{Broderick2016} found for Sgr~A*, the uncertainty in the inclination is quite large. Furthermore, \citet{GravityHS} found that the inclination angle of the orbital plane of the hotspot motion was $\sim 30^\circ$ during a flare. \autoref{fig:movie_cos} illustrates how the inclination angle changes the morphology of an image. As the inclination angle decreases, the impact of the lensed emission from the hotspot is suppressed since the disk becomes optically thick. Additionally, the variability of the light curve becomes subdued since the hotspot doesn't ``disappear'' behind the black hole at $\Theta=30^\circ,8^\circ$.  

  To analyze how disk inclination impacts hotspot measurements, we again followed the same procedure as above. That is, we created a twelve frame movie with scattering and background RIAF, and used \texttt{eht-imaging} to create a simulated dataset with the same array configuration as the previous experiments. The results of the parameter estimation are shown in \autoref{fig:triangle_cos}. \autoref{fig:triangle_cos} demonstrates that the inclination angle has a negligible effect on the ability of the EHT to extract hotspots when compared to the $\Theta=60^\circ$, again recovering spin to sub-percent precision.

\subsubsection{Future systematics to consider}
We have shown that the systematics included in this paper does not seem to impact the ability of the EHT to recover hotspot parameters.  However, other systematics need to be considered in future work. These include gain errors, refractive scattering, and variable background effects. 

Even after array calibration, it is expected that there will be residual $10\%$ gain errors in observations \citep{EHTCIII, ThemisCode}. In \citet{ThemisCode}, a gain mitigation technique was developed that was able to marginalize the impact of gains on parameter estimation and was applied in \citet{EHTCVI} and typically increased posterior width by a factor of a few.  Extrapolating from \cite{EHTCVI}, we do not expect that gains will then provide a significant obstacle to hotspot reconstruction.

While we have included diffractive scattering in this paper, Sgr~A* is also refractively scattered \citep{Bower2006, Johnson2018}. Refractive scattering effectively adds small scale structure to the movie impacting long-baseline visibilities.  However, we don't expect this to form a barrier to hotspot reconstruction for two reasons.    One, the timescale of the spot evolution is much shorter than the dynamical time scale of the scattering screen. The scattering timescale set by the orbital motion of the earth around the galactic center and is over hours, while hotspot changes on the order of minutes. Therefore, we can effectively treat the scattering screen as static during a flare. Second, as \autoref{fig:spot_ref} demonstrates, the scale of the scattering scintillation is typically on much smaller scales than hotspots.  This is due to diffractive scattering, which smears the hotspot to scales much larger than the refractive scintillation. Taken together, this suggests that while scattering mitigation is important, it should not significantly alter the results presented in this section.

Sgr A* displays consistent small-scale variability \citep{Witzel2018}, which is presumed to arise from turbulence and shocks in the accretion disk. General relativistic magnetohydrodynamics (GRMHD) simulations suggest that we consistently expect small-scale fluctuations in the accretion disk. For bright flares, this becomes less significant as a single region presumably dominates the emission. Nevertheless, to model the impact of this, we could include numerous sub-dominant spots, to model GRMHD turbulence, and then only attempt to recover the bright flare. Additionally, we could inject our hotspot model into a GRMHD simulation and then attempt to recover it\footnote{GRMHD simulation struggle to produce these flares since they typically ignore the microphysics and plasma resistivity needed to produce fast reconnection events.}.

\begin{figure*}[!ht]
\centering
\includegraphics[width=0.99\textwidth]{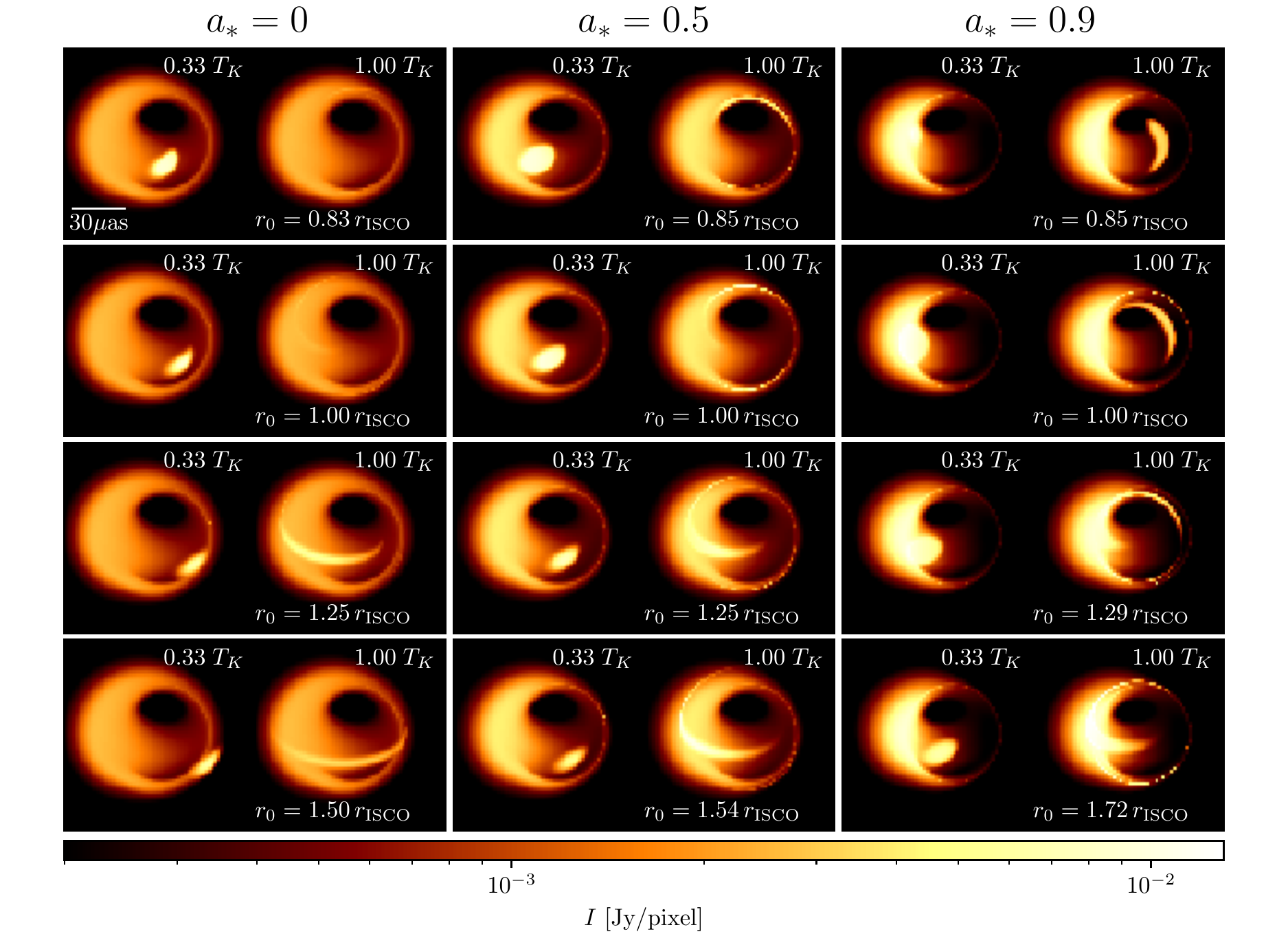}
\caption{Impact of changing spin parameter (increasing from left to right) and initial radius (increasing from top to bottom) of a shearing hotspot with a RIAF background, with colors in log-scale, showing intensity per pixel. In the top figure each row has a different with values $\mathrm{spin}=0,0.5,0.9$ from top to bottom. The bottom figure shows a different initial radius with values $r_0$ taken from Table~\ref{table:spot_experiment}, the spin for each spot is $0.5$.}\label{fig:spot_experiment_joint}
\end{figure*}

\begin{deluxetable*}{ccc|ccc|ccc}[!ht]\label{table:spot_experiment}
  \tablecaption{Spacetime \& Spot Parameters}
  \tablehead{
    \multicolumn3c{spin $0$} & \multicolumn3c{spin $0.5$} & \multicolumn3c{spin $0.9$}\\
    \colhead{$r_0/\risco$ } &\colhead{$r_0~[M]$ } & \colhead{Time [\rm hr]}&
    \colhead{$r_0/\risco$ } &\colhead{$r_0~[M]$ } & \colhead{Time [\rm hr]}&
    \colhead{$r_0/\risco$ } &\colhead{$r_0~[M]$ } & \colhead{Time [\rm hr]} 
  }
\startdata
 $0.85$   & $5.00 M$ & 0.83 & $0.85$ & $3.60 M$ & 0.54 & $0.85$   & $1.98 M$ & 0.27\\
 $1.0$    & $6.00 M$ & 1.1  & $1.0$  & $4.23 M$ & 0.68 & $1.0 $   & $2.33 M$ & 0.33\\
 $1.25$   & $7.50 M$ & 1.5  & $1.25$ & $5.30 M$ & 0.94 & $1.3$    & $3.00 M$ & 0.45\\
 $1.5$    & $9.00 M$ & 2.0  & $1.54$ & $6.50 M$ & 1.2  & $1.7$    & $4.00 M$ & 0.66\\
\enddata
\end{deluxetable*}
\hspace{1in}
\section{Spacetime tomography}\label{sec:tomo}

The frequency of flaring states in Sgr A* depends on the wavelength of the observations. At NIR, Sgr A* has a significant flare $\sim 4$ times per day \citep{Genzel2003, Eckart2006Pol,Meyer2009, Meyer2014, Hora2014}, while only a quarter of those typically have an X-ray counter part \citep{Baganoff2001,Eckart2004, Marrone2008,Porquet2008,Do2009,Neilsen2013, Mossoux2015}. Sub-mm occur 1--4 times a day (Marrone2008, Dexter2014). For the EHT however, it is not entirely clear whether the NIR/X-ray or sub-mm rate is relevant, given that the observations are at horizon scales. Either way, for any of the flaring rates of Sgr A*, the EHT will likely capture at least one flare per observational cycle. This implies that the EHT will measure multiple flares in the next few years.

These flares permit the opportunity to reconstruct the spacetime parameters in a position-dependent fashion, e.g., map the spacetime as a function of the initial hotspot radius.  The bundle of light rays (i.e., null geodesics) connecting the primary and higher-order images of a given hotspot will pass through different regions of the underlying spacetime for spots launched at different orbital radii.  Thus, the black hole mass and spin measurements from subsequent flaring epochs provide a spatially resolved probe of the black hole spacetime.  Such a spatially-resolved spacetime probe, or tomographical map of spacetime, provides a natural test of the no-hair theorem.  We will explore the limits that can be placed in practice on parameterized deviations from GR in a future publication.  One caveat to note is that in this paper we have chosen a typical model for these flares where the initial spot is in the disk. It is possible that the hotspot could form out of the plain of the disk or have significantly different accretion dynamics from those assumed in this paper. While this does mean our tomographical map of spacetime is model dependent, is provides an additional avenue to probe spacetime on event horizon scales. 

\subsection{Constructing a synthetic tomographical map of spacetime}
To explore the ability of EHT to perform spacetime tomography, we placed a series of hotspots in a Kerr spacetime varying both the initial radius of the hotspot and the spin of the black hole. Table~\ref{table:spot_experiment}, lists the radii and spins that were considered. As the initial radius, $r_0$, changes, the orbital period varies as well. To ensure that each experiment contains the same hotspot evolution, we restrict all movies to be $2 T_K(a_*,r_0)$ each using 12 frames.  Additionally, due to optical depth effects, the brightness of a hotspot will change as the spin and initial radius varies.  Therefore, for each movie, we picked the hotspot density normalization, $n_0$, such that the brightness was $\sim 0.5\, \rm Jy$'s above the quiescent emission. This ensures our results aren't due to the brightness of the hotspot. The other parameters, $\Theta,\  R_s,\  t_0, \phi_0,\  \alpha, \  \kappa,  \xi$, were held fixed between experiments and set to the same values in \autoref{sec:param_estimation}. Finally, each movie includes a background RIAF and diffractive scattering.

Figure \ref{fig:spot_experiment_joint}, shows how the intensity maps of each movie in Table~\ref{table:spot_experiment}, with the considerations in the above paragraph. The same subset of frames, in terms of $T_K$, is chosen for each movie. As the initial radius of the spot and spin of the black hole change so does that image by significant amounts. Furthermore, for small $r_0$ it becomes difficult to see the hotspot after one orbit due to the optical depth of the accretion disk. As we will see below, this can impact the ability of the EHT to recover hotspots close to the black hole.

\begin{figure}[!t]
\includegraphics[width=\linewidth]{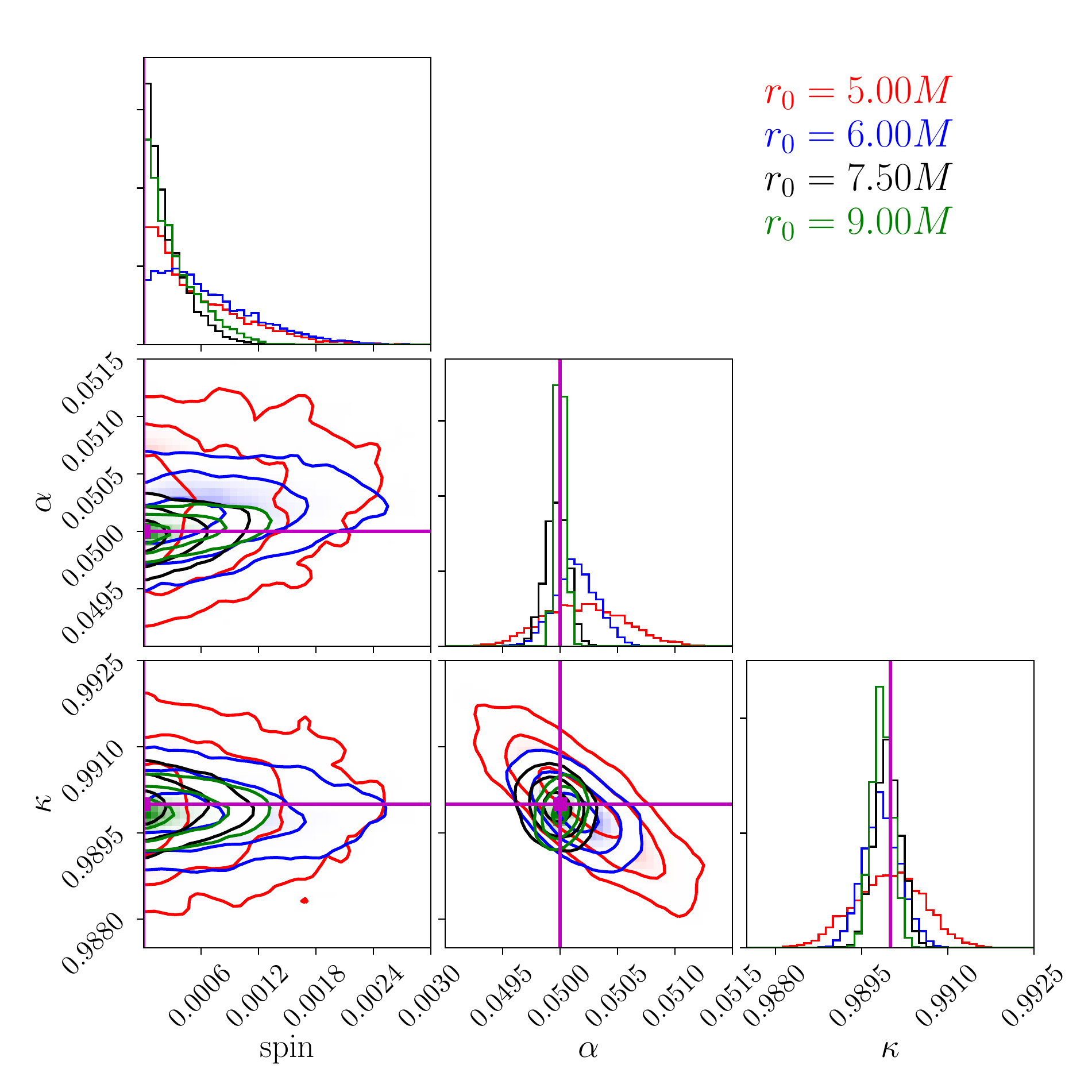}
\caption{Joint probability distributions of the spin $a_*$, and two accretion flow parameters $\alpha,\kappa$, for the experiments shown for $a_*=0$ and the radii $5M,6M,7.5M,9M$.}\label{fig:spot_experiment_joint_spin0}
\end{figure}
\begin{figure}
\includegraphics[width=\linewidth]{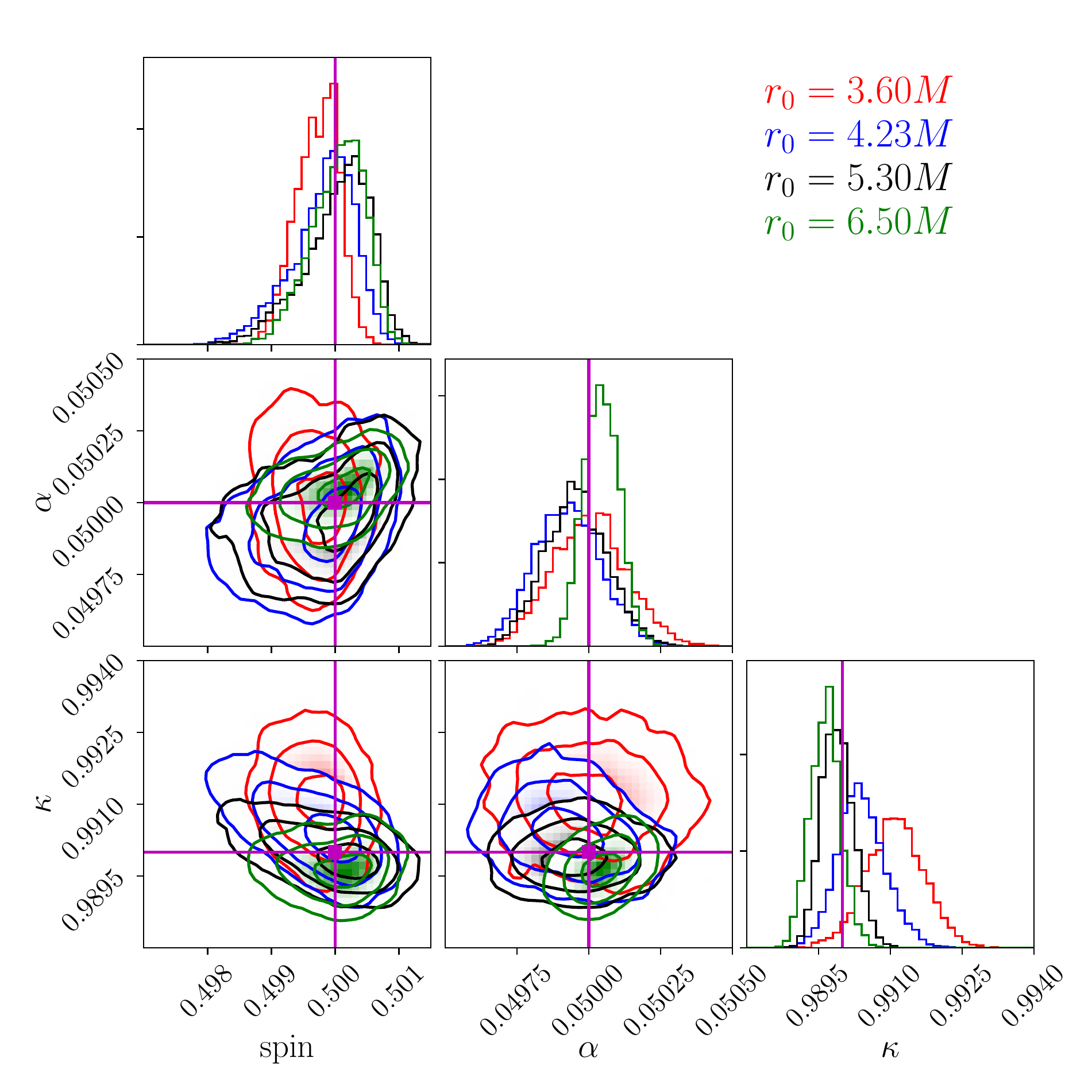}
\caption{Joint probability distributions of the spin $a_*$, and two accretion flow parameters $\alpha,\kappa$, for the experiments shown for $a_*=0.5$ and the radii $3.6M,4.23M,5.3M,6.5M$.}\label{fig:spot_experiment_joint_spin5}
\end{figure}
\begin{figure}
\includegraphics[width=\linewidth]{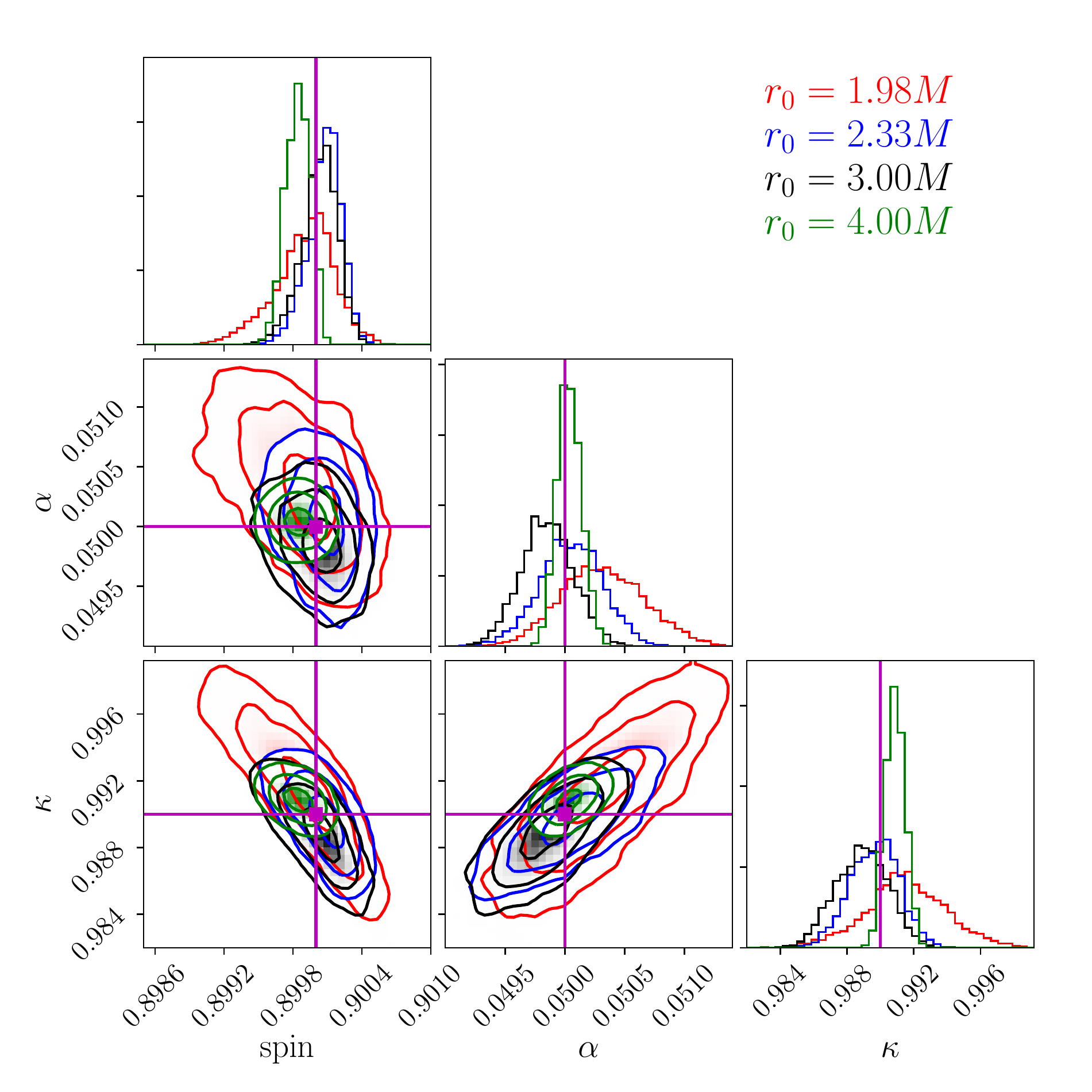}
\caption{Joint probability distributions of the spin $a_*$, and two accretion flow parameters $\alpha,\kappa$, for the experiments shown for $a_*=0.9$ and the radii $1.98M,2.33M,3M,4M$.}\label{fig:spot_experiment_joint_spin9}
\end{figure}

To create the synthetic EHT data, we used the same procedure described in \autoref{sec:ehtdata}. Namely, we used the same scan and integration time, observation frequency and bandwidth, and the same MJD and start time of the observation. Due to the movies being different lengths, each observation will have a different number of data points. This is a realistic simulation of actual spot observations since the duration of a flare sets the time interval we are interested in modeling.

\subsection{Results}
For parameter estimation, we used to same procedure described in Section~\ref{sec:param_estimation} and \ref{subsec:systematics}. The joint probability distributions for each run are shown in Figures \ref{fig:spot_experiment_joint_spin0}, \ref{fig:spot_experiment_joint_spin5}, and \ref{fig:spot_experiment_joint_spin9} for the spin $0,0.5,0.9$ cases respectively. Every experiment was able to recover the spot parameters to sub-percent levels using $95\%$ confidence levels about the median of the marginalized posteriors. Furthermore, the peaks of the joint-probability distributions are all statistically consistent with the true values of the model, which is shown by the purple line. 

Contrary to naive expectations based on spacetime considerations alone, the estimation of the black hole spin does not substantially improve as the spot moves closer to the black hole. This is because the optical depth from the accretion disk suppresses the intensity of the spot dramatically as it falls past the ISCO. For spots farther out, more of the hotspot is visible, giving much better constraints on the black hole spin.

\begin{figure}[!t]
\centering
\includegraphics[width=\linewidth]{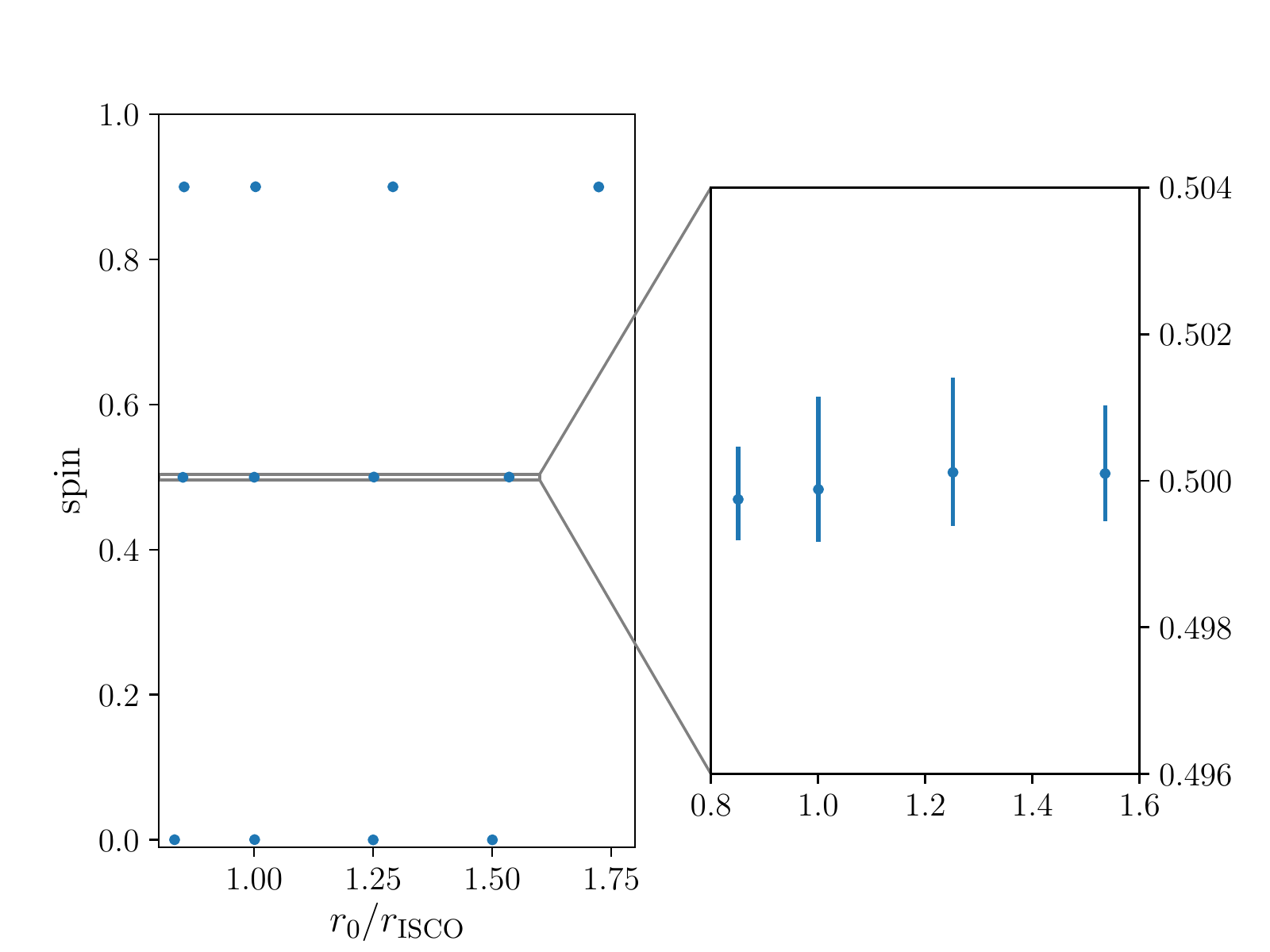}
\caption{Plots of recovered spin and initial radii parameters of shearing spots, with the effects of diffractive scattering and a background RIAF included.}
\label{fig:quantiles}
\end{figure}

Fixing the spin, we can associate each flare with a characteristic radius, i.e., the initial radius\footnote{While other choices are possible, the initial radius is the simplest and is a model parameter}.  The results of each column in Table~\ref{table:spot_experiment} forms a tomographical map of the given spacetime and is shown in \autoref{fig:quantiles}.  General relativity predicts that for a given spin, each flare must lie on a horizontal line in \autoref{fig:quantiles}. If there was evidence of curvature,  either the astrophysical model, e.g., accretion flow dynamics, is incorrect or that nature may deviate from GR near horizon scales. 

In summary, we have found that the EHT can tomographically map spacetime and accretion flow dynamics near the event horizon, providing a new test of GR in the strong gravity regime. While this test is not truly model-independent, it does provide an additional avenue to test the no-hair theorem, independent of others such as the black hole shadow size \citep{Johannsen2016,Psaltis2015,Psaltis2016}.

\hspace{1in}
\section{Conclusions}
Resolving structural variability on timescales of minutes to hours presents an opportunity to probe accretion processes and gravity on horizon scales. This is especially true for Sgr~A* that displays dramatic flaring events every 1--3 days. \cite{GravityHS} associated these flares with hotspots in the accretion disk thought to have arisen from magnetic reconnection events in the accretion disk and predicted over a decade ago \citet{BroderickLoeb2005,BroderickLoeb2006}.  It is expected that these spots will expand and shear as they traverse around the black hole since they are embedded in an accretion flow.  Therefore, we introduced a novel semi-analytical shearing hotspot model enabling us to perform parameter estimation studies with the EHT. 

Using the said model, we have shown that the 2017 EHT array can recover the hotspot parameters, such as spin, to sub-percent precision. Without any systematics, we were able to recover the spin to $~0.1\%$. Including diffractive scattering and a background accretion flow (see \autoref{fig:spot_riaf_sys}) we can recover spin to $~0.4\%$, and were able to show that this results did not depend on the black hole inclination. Furthermore, we were able to recover the spin to $0.05\%-0.5\%$ for variety of different initial radii as can be seen in Figures~\ref{fig:spot_experiment_joint_spin0}--\ref{fig:spot_experiment_joint_spin9}. By combining each of these results, we have demonstrated how observing hotspots naturally leads to a notion of mapping out the radial structure of accretion and spacetime at horizon scales and forming a tomographical map of spacetime. 

In future works, we plan to analyze how additional systematics impact the results in this paper.  Additionally, hotspots are the not only source of variability in Sgr A*. Turbulence and shocks in the accretion disk are thought to be responsible for most of the small-scale variability seen in Sgr A*. Therefore, we plan on analyzing the impact of such variability on hotspot parameter estimation. Related to this, is exploring whether multiple hotspots could be used to model the turbulence and shocks in GRMHD simulations and their impact on the results in this paper.

\acknowledgments
This work was made possible by the facilities of the Shared Hierarchical Academic Research Computing Network (SHARCNET:www.sharcnet.ca) and Compute/Calcul Canada (www.computecanada.ca).
Computations were made on the supercomputer Mammouth Parall\`ele 2 from University of Sherbrooke, managed by Calcul Qu\'ebec and Compute Canada. The operation of this supercomputer is funded by the Canada Foundation for Innovation (CFI), the minist\`ere de l'\'Economie, de la science et de l'innovation du Qu\'ebec (MESI) and the Fonds de recherche du Qu\'ebec - Nature et technologies (FRQ-NT).
This work was supported in part by Perimeter Institute for Theoretical Physics.  Research at Perimeter Institute is supported by the Government of Canada through the Department of Innovation, Science and Economic Development Canada and by the Province of Ontario through the Ministry of Economic Development, Job Creation and Trade.
P.T., A.E.B., and M.K. receive additional financial support from the Natural Sciences and Engineering Research Council of Canada through a Discovery Grant and additionally for PT through the Alexander Graham Bell Canada Graduate Scholarship (CGS-D). 
A.E.B. thanks the Delaney Family for their generous financial support via the Delaney Family John A. Wheeler Chair at Perimeter Institute.
R.G receives additional support from the ERC synergy grant “BlackHoleCam: Imaging the Event Horizon of Black Holes” (Grant No. 610058).

\bibliography{references.bib}

\begin{thebibliography}{}
\expandafter\ifx\csname natexlab\endcsname\relax\def\natexlab#1{#1}\fi
\providecommand{\url}[1]{\href{#1}{#1}}
\providecommand{\dodoi}[1]{doi:~\href{http://doi.org/#1}{\nolinkurl{#1}}}
\providecommand{\doeprint}[1]{\href{http://ascl.net/#1}{\nolinkurl{http://ascl.net/#1}}}
\providecommand{\doarXiv}[1]{\href{https://arxiv.org/abs/#1}{\nolinkurl{https://arxiv.org/abs/#1}}}

\bibitem[{{Akiyama} {et~al.}(2017){Akiyama}, {Kuramochi}, {Ikeda}, {Fish},
  {Tazaki}, {Honma}, {Doeleman}, {Broderick}, {Dexter}, {Mo{\'s}cibrodzka},
  {Bouman}, {Chael}, \& {Zaizen}}]{Akiyama2017}
{Akiyama}, K., {Kuramochi}, K., {Ikeda}, S., {et~al.} 2017, \apj, 838, 1,
  \dodoi{10.3847/1538-4357/aa6305}

\bibitem[{{Baganoff} {et~al.}(2001){Baganoff}, {Bautz}, {Brandt}, {Chartas},
  {Feigelson}, {Garmire}, {Maeda}, {Morris}, {Ricker}, {Townsley}, \&
  {Walter}}]{Baganoff2001}
{Baganoff}, F.~K., {Bautz}, M.~W., {Brandt}, W.~N., {et~al.} 2001, \nat, 413,
  45, \dodoi{10.1038/35092510}

\bibitem[{{Bower} {et~al.}(2006){Bower}, {Goss}, {Falcke}, {Backer}, \&
  {Lithwick}}]{Bower2006}
{Bower}, G.~C., {Goss}, W.~M., {Falcke}, H., {Backer}, D.~C., \& {Lithwick}, Y.
  2006, \apjl, 648, L127, \dodoi{10.1086/508019}

\bibitem[{{Broderick} \& {Blandford}(2004)}]{Broderick2004}
{Broderick}, A., \& {Blandford}, R. 2004, \mnras, 349, 994,
  \dodoi{10.1111/j.1365-2966.2004.07582.x}

\bibitem[{Broderick {et~al.}(2011)Broderick, Fish, Doeleman, \&
  Loeb}]{Broderick2011}
Broderick, A.~E., Fish, V.~L., Doeleman, S.~S., \& Loeb, A. 2011, The
  Astrophysical Journal, 735, 110

\bibitem[{Broderick {et~al.}(in prep.)Broderick, Gold, Karami, Preciado-Lopez,
  Tiede, \& Pu}]{ThemisCode}
Broderick, A.~E., Gold, R., Karami, M., {et~al.} in prep.

\bibitem[{Broderick {et~al.}(2014)Broderick, Johannsen, Loeb, \&
  Psaltis}]{Broderick2014}
Broderick, A.~E., Johannsen, T., Loeb, A., \& Psaltis, D. 2014, The
  Astrophysical Journal, 784, 7

\bibitem[{{Broderick} \& {Loeb}(2005)}]{BroderickLoeb2005}
{Broderick}, A.~E., \& {Loeb}, A. 2005, \mnras, 363, 353,
  \dodoi{10.1111/j.1365-2966.2005.09458.x}

\bibitem[{{Broderick} \& {Loeb}(2006)}]{BroderickLoeb2006}
---. 2006, \mnras, 367, 905, \dodoi{10.1111/j.1365-2966.2006.10152.x}

\bibitem[{Broderick {et~al.}(2016)Broderick, Fish, Johnson, Rosenfeld, Wang,
  Doeleman, Akiyama, Johannsen, \& Roy}]{Broderick2016}
Broderick, A.~E., Fish, V.~L., Johnson, M.~D., {et~al.} 2016, \apj, 820, 137

\bibitem[{Chael {et~al.}(2019)Chael, Bouman, Johnson, Wielgus, Blackburn, kwan
  Chan, Farah, Palumbo, \& Pesce}]{ehtim2019}
Chael, A., Bouman, K., Johnson, M., {et~al.} 2019, {eht-imaging: v1.1.0:
  Imaging interferometric data with regularized maximum likelihood},
  \dodoi{10.5281/zenodo.2614016}.
\newblock \url{https://doi.org/10.5281/zenodo.2614016}

\bibitem[{{Chael} {et~al.}(2018){Chael}, {Johnson}, {Bouman}, {Blackburn},
  {Akiyama}, \& {Narayan}}]{ehtim2018}
{Chael}, A.~A., {Johnson}, M.~D., {Bouman}, K.~L., {et~al.} 2018, \apj, 857,
  23, \dodoi{10.3847/1538-4357/aab6a8}

\bibitem[{{Chael} {et~al.}(2016){Chael}, {Johnson}, {Narayan}, {Doeleman},
  {Wardle}, \& {Bouman}}]{ehtim2016}
{Chael}, A.~A., {Johnson}, M.~D., {Narayan}, R., {et~al.} 2016, \apj, 829, 11,
  \dodoi{10.3847/0004-637X/829/1/11}

\bibitem[{{Chan} {et~al.}(2015){Chan}, {Psaltis}, {{\"O}zel}, {Narayan}, \&
  {Sa{\c d}owski}}]{Chan2015}
{Chan}, C.-K., {Psaltis}, D., {{\"O}zel}, F., {Narayan}, R., \& {Sa{\c
  d}owski}, A. 2015, \apj, 799, 1, \dodoi{10.1088/0004-637X/799/1/1}

\bibitem[{{Do} {et~al.}(2009){Do}, {Ghez}, {Morris}, {Yelda}, {Meyer}, {Lu},
  {Hornstein}, \& {Matthews}}]{Do2009}
{Do}, T., {Ghez}, A.~M., {Morris}, M.~R., {et~al.} 2009, \apj, 691, 1021,
  \dodoi{10.1088/0004-637X/691/2/1021}

\bibitem[{{Doeleman} {et~al.}(2009){Doeleman}, {Agol}, {Backer}, {Baganoff},
  {Bower}, {Broderick}, {Fabian}, {Fish}, {Gammie}, {Ho}, {Honman},
  {Krichbaum}, {Loeb}, {Marrone}, {Reid}, {Rogers}, {Shapiro}, {Strittmatter},
  {Tilanus}, {Weintroub}, {Whitney}, {Wright}, \&
  {Ziurys}}]{2009astro2010S..68D}
{Doeleman}, S., {Agol}, E., {Backer}, D., {et~al.} 2009, in ArXiv Astrophysics
  e-prints, Vol. 2010, astro2010: The Astronomy and Astrophysics Decadal
  Survey, 68

\bibitem[{Doeleman {et~al.}(2008)Doeleman, Weintroub, Rogers,
  {et~al.}}]{Doeleman2008}
Doeleman, S.~S., Weintroub, J., Rogers, A.~E.~E., {et~al.} 2008, \nat, 455, 78,
  \dodoi{10.1038/nature07245}

\bibitem[{{Dov{\v{c}}iak} {et~al.}(2004){Dov{\v{c}}iak}, {Karas}, \&
  {Yaqoob}}]{Dovciak2004}
{Dov{\v{c}}iak}, M., {Karas}, V., \& {Yaqoob}, T. 2004, \apjs, 153, 205,
  \dodoi{10.1086/421115}

\bibitem[{{Eckart} {et~al.}(2006){Eckart}, {Sch{\"o}del}, {Meyer}, {Trippe},
  {Ott}, \& {Genzel}}]{Eckart2006Pol}
{Eckart}, A., {Sch{\"o}del}, R., {Meyer}, L., {et~al.} 2006, \aap, 455, 1,
  \dodoi{10.1051/0004-6361:20064948}

\bibitem[{{Eckart} {et~al.}(2004){Eckart}, {Baganoff}, {Morris}, {Bautz},
  {Brandt}, {Garmire}, {Genzel}, {Ott}, {Ricker}, {Straubmeier}, {Viehmann},
  {Sch{\"o}del}, {Bower}, \& {Goldston}}]{Eckart2004}
{Eckart}, A., {Baganoff}, F.~K., {Morris}, M., {et~al.} 2004, \aap, 427, 1,
  \dodoi{10.1051/0004-6361:20040495}

\bibitem[{{Eckart} {et~al.}(2008){Eckart}, {Baganoff}, {Zamaninasab}, {Morris},
  {Sch{\"o}del}, {Meyer}, {Muzic}, {Bautz}, {Brandt}, {Garmire}, {Ricker},
  {Kunneriath}, {Straubmeier}, {Duschl}, {Dovciak}, {Karas}, {Markoff},
  {Najarro}, {Mauerhan}, {Moultaka}, \& {Zensus}}]{Eckart2008a}
{Eckart}, A., {Baganoff}, F.~K., {Zamaninasab}, M., {et~al.} 2008, \aap, 479,
  625, \dodoi{10.1051/0004-6361:20078793}

\bibitem[{{Eckart} {et~al.}(2009){Eckart}, {Baganoff}, {Morris}, {Kunneriath},
  {Zamaninasab}, {Witzel}, {Sch{\"o}del}, {Garc{\'{\i}}a-Mar{\'{\i}}n},
  {Meyer}, {Bower}, {Marrone}, {Bautz}, {Brandt}, {Garmire}, {Ricker},
  {Straubmeier}, {Roberts}, {Muzic}, {Mauerhan}, \& {Zensus}}]{Eckart2009}
{Eckart}, A., {Baganoff}, F.~K., {Morris}, M.~R., {et~al.} 2009, \aap, 500,
  935, \dodoi{10.1051/0004-6361/200811354}

\bibitem[{{Event Horizon Telescope Collaboration}
  {et~al.}(2019{\natexlab{a}}){Event Horizon Telescope Collaboration},
  {Akiyama}, {Alberdi}, {Alef}, {Asada}, {Azulay}, {Baczko}, {Ball},
  {Balokovi{\'c}}, \& {Barrett}}]{EHTCI}
{Event Horizon Telescope Collaboration}, {Akiyama}, K., {Alberdi}, A., {et~al.}
  2019{\natexlab{a}}, \apj, 875, L1, \dodoi{10.3847/2041-8213/ab0ec7}

\bibitem[{{Event Horizon Telescope Collaboration}
  {et~al.}(2019{\natexlab{b}}){Event Horizon Telescope Collaboration},
  {Akiyama}, {Alberdi}, {Alef}, {Asada}, {Azulay}, {Baczko}, {Ball},
  {Balokovi{\'c}}, \& {Barrett}}]{EHTCII}
---. 2019{\natexlab{b}}, \apj, 875, L2, \dodoi{10.3847/2041-8213/ab0c96}

\bibitem[{{Event Horizon Telescope Collaboration}
  {et~al.}(2019{\natexlab{c}}){Event Horizon Telescope Collaboration},
  {Akiyama}, {Alberdi}, {Alef}, {Asada}, {Azulay}, {Baczko}, {Ball},
  {Balokovi{\'c}}, \& {Barrett}}]{EHTCIII}
---. 2019{\natexlab{c}}, \apj, 875, L3, \dodoi{10.3847/2041-8213/ab0c57}

\bibitem[{{Event Horizon Telescope Collaboration}
  {et~al.}(2019{\natexlab{d}}){Event Horizon Telescope Collaboration},
  {Akiyama}, {Alberdi}, {Alef}, {Asada}, {Azulay}, {Baczko}, {Ball},
  {Balokovi{\'c}}, \& {Barrett}}]{EHTCIV}
---. 2019{\natexlab{d}}, \apj, 875, L4, \dodoi{10.3847/2041-8213/ab0e85}

\bibitem[{{Event Horizon Telescope Collaboration}
  {et~al.}(2019{\natexlab{e}}){Event Horizon Telescope Collaboration},
  {Akiyama}, {Alberdi}, {Alef}, {Asada}, {Azulay}, {Baczko}, {Ball},
  {Balokovi{\'c}}, \& {Barrett}}]{EHTCV}
---. 2019{\natexlab{e}}, \apj, 875, L5, \dodoi{10.3847/2041-8213/ab0f43}

\bibitem[{{Event Horizon Telescope Collaboration}
  {et~al.}(2019{\natexlab{f}}){Event Horizon Telescope Collaboration},
  {Akiyama}, {Alberdi}, {Alef}, {Asada}, {Azulay}, {Baczko}, {Ball},
  {Balokovi{\'c}}, \& {Barrett}}]{EHTCVI}
---. 2019{\natexlab{f}}, \apj, 875, L6, \dodoi{10.3847/2041-8213/ab1141}

\bibitem[{{Fazio} {et~al.}(2018){Fazio}, {Hora}, {Witzel}, {Willner}, {Ashby},
  {Baganoff}, {Becklin}, {Carey}, {Haggard}, {Gammie}, {Ghez}, {Gurwell},
  {Ingalls}, {Marrone}, {Morris}, \& {Smith}}]{Fazio2018}
{Fazio}, G.~G., {Hora}, J.~L., {Witzel}, G., {et~al.} 2018, \apj, 864, 58,
  \dodoi{10.3847/1538-4357/aad4a2}

\bibitem[{Fish {et~al.}(2011)Fish, Doeleman, Beaudoin, Blundell, Bolin, Bower,
  Chamberlin, Freund, Friberg, Gurwell, Honma, Inoue, Krichbaum, Lamb, Marrone,
  Moran, Oyama, Plambeck, Primiani, Rogers, Smythe, SooHoo, Strittmatter,
  Tilanus, Titus, Weintroub, Wright, Woody, Young, \& Ziurys}]{Fish2011}
Fish, V.~L., Doeleman, S.~S., Beaudoin, C., {et~al.} 2011, \apjl, 727, L36,
  \dodoi{10.1088/2041-8205/727/2/L36}

\bibitem[{{Fish} {et~al.}(2016){Fish}, {Johnson}, {Doeleman}, {Broderick},
  {Psaltis}, {Lu}, {Akiyama}, {Alef}, {Algaba}, {Asada}, {Beaudoin},
  {Bertarini}, {Blackburn}, {Blundell}, {Bower}, {Brinkerink}, {Cappallo},
  {Chael}, {Chamberlin}, {Chan}, {Crew}, {Dexter}, {Dexter}, {Dzib}, {Falcke},
  {Freund}, {Friberg}, {Greer}, {Gurwell}, {Ho}, {Honma}, {Inoue}, {Johannsen},
  {Kim}, {Krichbaum}, {Lamb}, {Le{\'o}n-Tavares}, {Loeb}, {Loinard},
  {MacMahon}, {Marrone}, {Moran}, {Mo{\'s}cibrodzka}, {Ortiz-Le{\'o}n},
  {Oyama}, {{\"O}zel}, {Plambeck}, {Pradel}, {Primiani}, {Rogers}, {Rosenfeld},
  {Rottmann}, {Roy}, {Ruszczyk}, {Smythe}, {SooHoo}, {Spilker}, {Stone},
  {Strittmatter}, {Tilanus}, {Titus}, {Vertatschitsch}, {Wagner}, {Wardle},
  {Weintroub}, {Woody}, {Wright}, {Yamaguchi}, {Young}, {Young}, {Zensus}, \&
  {Ziurys}}]{Fish2016}
{Fish}, V.~L., {Johnson}, M.~D., {Doeleman}, S.~S., {et~al.} 2016, \apj, 820,
  90, \dodoi{10.3847/0004-637X/820/2/90}

\bibitem[{Genzel {et~al.}(2003)Genzel, Sch\"{o}del, Ott, \&
  Eckart}]{Genzel2003}
Genzel, R., Sch\"{o}del, R., Ott, T., \& Eckart, A. 2003, Nature, 425, 7

\bibitem[{{Gillessen} {et~al.}(2006){Gillessen}, {Eisenhauer}, {Quataert},
  {Genzel}, {Paumard}, {Trippe}, {Ott}, {Abuter}, {Eckart}, {Lagage},
  {Lehnert}, {Tacconi}, \& {Martins}}]{Gillessen2006}
{Gillessen}, S., {Eisenhauer}, F., {Quataert}, E., {et~al.} 2006, Journal of
  Physics Conference Series, 54, 411, \dodoi{10.1088/1742-6596/54/1/065}

\bibitem[{Goodman \& Weare(2010)}]{Goodman2010}
Goodman, J., \& Weare, J. 2010, Communications in Applied Mathematics and
  Computational Science, 5, 65, \dodoi{10.2140/camcos.2010.5.65}

\bibitem[{{Gravity Collaboration} {et~al.}(2018){Gravity Collaboration},
  {Abuter}, {Amorim}, {Baub{\"o}ck}, {Berger}, {Bonnet}, {Brand ner},
  {Cl{\'e}net}, {Coud{\'e} Du Foresto}, \& {de Zeeuw}}]{GravityHS}
{Gravity Collaboration}, {Abuter}, R., {Amorim}, A., {et~al.} 2018, \aap, 618,
  L10, \dodoi{10.1051/0004-6361/201834294}

\bibitem[{{Hora} {et~al.}(2014){Hora}, {Witzel}, {Ashby}, {Becklin}, {Carey},
  {Fazio}, {Ghez}, {Ingalls}, {Meyer}, {Morris}, {Smith}, \&
  {Willner}}]{Hora2014}
{Hora}, J.~L., {Witzel}, G., {Ashby}, M.~L.~N., {et~al.} 2014, \apj, 793, 120,
  \dodoi{10.1088/0004-637X/793/2/120}

\bibitem[{{Issaoun} {et~al.}(2019){Issaoun}, {Johnson}, {Blackburn},
  {Brinkerink}, {Mo{\'s}cibrodzka}, {Chael}, {Goddi}, {Mart{\'\i}-Vidal},
  {Wagner}, \& {Doeleman}}]{Issaoun2019}
{Issaoun}, S., {Johnson}, M.~D., {Blackburn}, L., {et~al.} 2019, \apj, 871, 30,
  \dodoi{10.3847/1538-4357/aaf732}

\bibitem[{Johannsen {et~al.}(2016{\natexlab{a}})Johannsen, Wang, Broderick,
  Doeleman, Fish, Loeb, \& Psaltis}]{JohanssenII}
Johannsen, T., Wang, C., Broderick, A.~E., {et~al.} 2016{\natexlab{a}}, Phys.
  Rev. Lett., 117, 091101, \dodoi{10.1103/PhysRevLett.117.091101}

\bibitem[{Johannsen {et~al.}(2016{\natexlab{b}})Johannsen, Broderick, Plewa,
  Chatzopoulos, Doeleman, Eisenhauer, Fish, Genzel, Gerhard, \&
  Johnson}]{Johannsen2016}
Johannsen, T., Broderick, A.~E., Plewa, P.~M., {et~al.} 2016{\natexlab{b}},
  Phys. Rev. Lett., 116, 031101, \dodoi{10.1103/PhysRevLett.116.031101}

\bibitem[{Johnson {et~al.}(2015)Johnson, Fish, Doeleman, Marrone, Plambeck,
  Wardle, Akiyama, Asada, Beaudoin, Blackburn, Blundell, Bower, Brinkerink,
  Broderick, Cappallo, Chael, Crew, Dexter, Dexter, Freund, Friberg, Gold,
  Gurwell, Ho, Honma, Inoue, Kosowsky, Krichbaum, Lamb, Loeb, Lu, MacMahon,
  McKinney, Moran, Narayan, Primiani, Psaltis, Rogers, Rosenfeld, SooHoo,
  Tilanus, Titus, Vertatschitsch, Weintroub, Wright, Young, Zensus, \&
  Ziurys}]{Johnson:2015}
Johnson, M.~D., Fish, V.~L., Doeleman, S.~S., {et~al.} 2015, Science, 350,
  1242, \dodoi{10.1126/science.aac7087}

\bibitem[{{Johnson} {et~al.}(2018){Johnson}, {Narayan}, {Psaltis}, {Blackburn},
  {Kovalev}, {Gwinn}, {Zhao}, {Bower}, {Moran}, \& {Kino}}]{Johnson2018}
{Johnson}, M.~D., {Narayan}, R., {Psaltis}, D., {et~al.} 2018, \apj, 865, 104,
  \dodoi{10.3847/1538-4357/aadcff}

\bibitem[{{Marrone}(2006)}]{Marrone2006}
{Marrone}, D.~P. 2006, PhD thesis, Harvard University

\bibitem[{{Marrone} {et~al.}(2008){Marrone}, {Baganoff}, {Morris}, {Moran},
  {Ghez}, {Hornstein}, {Dowell}, {Mu{\~n}oz}, {Bautz}, {Ricker}, {Brandt},
  {Garmire}, {Lu}, {Matthews}, {Zhao}, {Rao}, \& {Bower}}]{Marrone2008}
{Marrone}, D.~P., {Baganoff}, F.~K., {Morris}, M.~R., {et~al.} 2008, \apj, 682,
  373, \dodoi{10.1086/588806}

\bibitem[{{Meyer} {et~al.}(2009){Meyer}, {Do}, {Ghez}, {Morris}, {Yelda},
  {Sch{\"o}del}, \& {Eckart}}]{Meyer2009}
{Meyer}, L., {Do}, T., {Ghez}, A., {et~al.} 2009, \apjl, 694, L87,
  \dodoi{10.1088/0004-637X/694/1/L87}

\bibitem[{{Meyer} {et~al.}(2014){Meyer}, {Witzel}, {Longstaff}, \&
  {Ghez}}]{Meyer2014}
{Meyer}, L., {Witzel}, G., {Longstaff}, F.~A., \& {Ghez}, A.~M. 2014, \apj,
  791, 24, \dodoi{10.1088/0004-637X/791/1/24}

\bibitem[{{Mossoux} {et~al.}(2015){Mossoux}, {Grosso}, {Vincent}, \&
  {Porquet}}]{Mossoux2015}
{Mossoux}, E., {Grosso}, N., {Vincent}, F.~H., \& {Porquet}, D. 2015, in
  SF2A-2015: Proceedings of the Annual meeting of the French Society of
  Astronomy and Astrophysics, 171--174

\bibitem[{{Neilsen} {et~al.}(2013){Neilsen}, {Nowak}, {Gammie}, {Dexter},
  {Markoff}, {Haggard}, {Nayakshin}, {Wang}, {Grosso}, {Porquet}, {Tomsick},
  {Degenaar}, {Fragile}, {Houck}, {Wijnands}, {Miller}, \&
  {Baganoff}}]{Neilsen2013}
{Neilsen}, J., {Nowak}, M.~A., {Gammie}, C., {et~al.} 2013, \apj, 774, 42,
  \dodoi{10.1088/0004-637X/774/1/42}

\bibitem[{{Ortiz-Le{\'o}n} {et~al.}(2016){Ortiz-Le{\'o}n}, {Johnson},
  {Doeleman}, {Blackburn}, {Fish}, {Loinard}, {Reid}, {Castillo}, {Chael},
  {Hern{\'a}ndez-G{\'o}mez}, {Hughes}, {Le{\'o}n-Tavares}, {Lu}, {Monta{\~n}a},
  {Narayanan}, {Rosenfeld}, {S{\'a}nchez}, {Schloerb}, {Shen}, {Shiokawa},
  {SooHoo}, \& {Vertatschitsch}}]{Ortiz-Le2016}
{Ortiz-Le{\'o}n}, G.~N., {Johnson}, M.~D., {Doeleman}, S.~S., {et~al.} 2016,
  \apj, 824, 40, \dodoi{10.3847/0004-637X/824/1/40}

\bibitem[{{Ponti} {et~al.}(2017){Ponti}, {George}, {Scaringi}, {Zhang}, {Jin},
  {Dexter}, {Terrier}, {Clavel}, {Degenaar}, {Eisenhauer}, {Genzel},
  {Gillessen}, {Goldwurm}, {Habibi}, {Haggard}, {Hailey}, {Harrison},
  {Merloni}, {Mori}, {Nandra}, {Ott}, {Pfuhl}, {Plewa}, \& {Waisberg}}]{Ponti}
{Ponti}, G., {George}, E., {Scaringi}, S., {et~al.} 2017, \mnras, 468, 2447,
  \dodoi{10.1093/mnras/stx596}

\bibitem[{{Porquet} {et~al.}(2008){Porquet}, {Grosso}, {Predehl}, {Hasinger},
  {Yusef-Zadeh}, {Aschenbach}, {Trap}, {Melia}, {Warwick}, {Goldwurm},
  {B{\'e}langer}, {Tanaka}, {Genzel}, {Dodds-Eden}, {Sakano}, \&
  {Ferrando}}]{Porquet2008}
{Porquet}, D., {Grosso}, N., {Predehl}, P., {et~al.} 2008, \aap, 488, 549,
  \dodoi{10.1051/0004-6361:200809986}

\bibitem[{{Psaltis} {et~al.}(2015){Psaltis}, {{\"O}zel}, {Chan}, \&
  {Marrone}}]{Psaltis2015}
{Psaltis}, D., {{\"O}zel}, F., {Chan}, C.-K., \& {Marrone}, D.~P. 2015, \apj,
  814, 115, \dodoi{10.1088/0004-637X/814/2/115}

\bibitem[{{Psaltis} {et~al.}(2016){Psaltis}, {Wex}, \& {Kramer}}]{Psaltis2016}
{Psaltis}, D., {Wex}, N., \& {Kramer}, M. 2016, \apj, 818, 121,
  \dodoi{10.3847/0004-637X/818/2/121}

\bibitem[{{Pu} {et~al.}(2016){Pu}, {Akiyama}, \& {Asada}}]{Pu2016}
{Pu}, H.-Y., {Akiyama}, K., \& {Asada}, K. 2016, \apj, 831, 4,
  \dodoi{10.3847/0004-637X/831/1/4}

\bibitem[{{Thompson} {et~al.}(2017){Thompson}, {Moran}, \& {Swenson}}]{TMS}
{Thompson}, A.~R., {Moran}, J.~M., \& {Swenson}, Jr., G.~W. 2017,
  {Interferometry and Synthesis in Radio Astronomy, 3rd Edition} ({Springer
  International Publishing}), \dodoi{10.1007/978-3-319-44431-4}

\bibitem[{{Vousden} {et~al.}(2016){Vousden}, {Farr}, \& {Mandel}}]{Vousden16}
{Vousden}, W.~D., {Farr}, W.~M., \& {Mandel}, I. 2016, \mnras, 455, 1919,
  \dodoi{10.1093/mnras/stv2422}

\bibitem[{Witzel {et~al.}(2012)Witzel, Eckart, Bremer, Zamaninasab,
  Shahzamanian, Valencia-S., Sch\"{o}del, Karas, Lenzen, Marchili, Sabha,
  Garcia-Marin, Buchholz, Kunneriath, \& Straubmeier}]{Witzel2012}
Witzel, G., Eckart, a., Bremer, M., {et~al.} 2012, ApJ Suppl. Series, 203, 18,
  \dodoi{10.1088/0067-0049/203/2/18}

\bibitem[{{Witzel} {et~al.}(2018){Witzel}, {Martinez}, {Hora}, {Willner},
  {Morris}, {Gammie}, {Becklin}, {Ashby}, {Baganoff}, {Carey}, {Do}, {Fazio},
  {Ghez}, {Glaccum}, {Haggard}, {Herrero-Illana}, {Ingalls}, {Narayan}, \&
  {Smith}}]{Witzel2018}
{Witzel}, G., {Martinez}, G., {Hora}, J., {et~al.} 2018, \apj, 863, 15,
  \dodoi{10.3847/1538-4357/aace62}

\bibitem[{{Zamaninasab} {et~al.}(2010){Zamaninasab}, {Eckart}, {Witzel},
  {Dovciak}, {Karas}, {Sch{\"o}del}, {Gie{\ss}{\"u}bel}, {Bremer},
  {Garc{\'{\i}}a-Mar{\'{\i}}n}, {Kunneriath}, {Mu{\v z}i{\'c}}, {Nishiyama},
  {Sabha}, {Straubmeier}, \& {Zensus}}]{ZamaninasabShear}
{Zamaninasab}, M., {Eckart}, A., {Witzel}, G., {et~al.} 2010, \aap, 510, A3,
  \dodoi{10.1051/0004-6361/200912473}

\end{thebibliography}
\end{document}